\documentclass[12pt,a4paper]{article} 

\usepackage{latexsym}
\usepackage{epsf}
\usepackage{graphicx}

\def\be{\begin{equation}}
\def\ee{\end{equation}}
\def\bea{\begin{eqnarray}}
\def\eea{\end{eqnarray}}
\def\by{\left(\begin{array}}
\def\ey{\end{array}\right)}

\def\slash#1{\setbox0=\hbox{$#1$}#1\hskip-\wd0\dimen0=5pt\advance
       \dimen0 by-\ht0\advance\dimen0 by\dp0\lower0.5\dimen0\hbox
         to\wd0{\hss\sl/\/\hss}}

\newcommand{\sla}[1]{\mbox{$#1\!\!\!/$}}

\begin{document}
\begin{center}
\Large{Renormalization of the two--photon vacuum polarization }\\
\Large{and the self energy vacuum polarization}\\
\Large{for a tightly bound electron}
\end{center}
\begin{center}

\small {Sven Zschocke, G\"unter Plunien, and Gerhard Soff}
\end{center}

\footnotesize{
\begin{center}
Institut f\"ur Theoretische Physik \\
Technische Universit\"at Dresden, Mommsenstrasse 13 \\
D--01062 Dresden, Germany
\end{center}

\normalsize
\date{\today}
\begin{abstract}
The renormalization method of Bogoljubov--Parasiuk--Hepp--Zimmermann (BPHZ) is 
used in order to derive the renormalized energy shift due to the gauge 
invariant K\"all\'en--Sabry diagram of the two--photon vacuum polarization 
(VPVP) as well as the self energy vacuum polarization S(VP)E beyond the Uehling 
approximation. It is outlined, that no outer renormalization is required for the 
two--photon vacuum polarization and that only the inner renormalization has to be 
accomplished. It is shown that the so--called nongauge invariant spurious 
term is absent for a wide class of vacuum polarization (VP) diagrams if one 
applies the widely used spherical expansion of bound and free--electron propagator. 
This simplifies significantly calculations in bound state quantum 
electrodynamics. 
As one result of our paper the use of the BPHZ--approach in bound state QED 
is established.
\end{abstract}
\section{Introduction}
Highly charged ions provide an ideal scenario to demonstrate the validity
of QED in strong electric and magnetic fields by measurements of the Lamb 
shift at utmost precision. In this respect the recent experimental progress 
made in measurements of the Lamb shift in hydrogenlike ions 
\cite{intro0,intro1} indicate that calculations of all QED corrections of 
order $\alpha^2$ ($\alpha\approx 1/137.036$ is the fine structure constant), 
but exact in the coupling constant 
$Z \alpha$ become relevant. The present 
status of theoretical prediction for the Lamb shift have been presented in 
\cite{intro2,intro3}. 
Even the most difficult gauge invariant set, the second--order 
electron self energy correction, have been recently calculated for the  
hydrogenlike ions uranium and bismuth \cite{intro4}. 
The renormalization of these diagrams has been carried out in Ref. 
\cite{intro5} and within a different approach in Ref. \cite{intro6}.
At present two of these two--photon diagrams are known in the Uehling 
approximation only, namely the two--photon vacuum polarization VPVP and the 
effective self energy S(VP)E (see Fig. \ref{fig0}). Concequently, mainly the 
VPVP correction remains an additional source of  
theoretical uncertainty in the Lamb shift predictions of hydrogenlike ions. 
\cite{intro2,intro3}.  
The main contribution originates from the Uehling 
part of the VPVP diagram  
($-0.6 \pm 0.12$ eV for uranium and $-0.34 \pm 0.07$ eV for lead). 
To provide a result, complete in order $\alpha^2$, for the Lamb shift 
is indeed a challenge for 
theory. As a step towards this goal we derive the 
renormalized expression for the two--photon vacuum polarization. 
At the same time we derive a renormalized expression for the 
higher--order energy shift of the effective self energy diagram S(VP)E.
This diagram is of special interest not only since its contribution 
to the Lamb shift is unknown for all elements in nature but also due  
to the fact that it contains the one--loop vacuum polarization (VP) as a 
subdiagram. It is 
well--known that in general the evaluation of VP is connected with a so--called 
nongauge invariant spurious term. We shall show that this spurious term is 
absent for a wide class of diagrams which contain the VP diagram.
This result simplifies essentially evaluations of such diagrams in 
bound state QED.\\ 
The paper is organized as follows. In section {\tt 2} we explain the 
BPHZ--renormalization approach and it's application to bound state QED.
The renormalization of the VPVP diagram is discussed in section {\tt 3}. 
It is shown that no outer renormalization is required and the 
renormalized expression is specified in section {\tt 4}. 
In section {\tt 5} we shall discuss some basic properties of the fourth--rank vacuum 
polarization tensor. In section {\tt 6} the renormalized expression of higher--order 
of the S(VP)E diagram is derived. Finally it will be demonstrated that the 
nongauge invariant term is absent if one applies the  
partial wave decomposition of the electron propagator. The renormalized 
expression of S(VP)E diagram is specified in section {\tt 7}.

\section{BPHZ--Renormalization scheme for QED}
The general renormalization scheme of QED is formulated for graphs with 
free--electron propagators. Therefore, in order to isolate the ultraviolet 
divergencies we first have to perform a potential expansion \cite{section1_1, 
section1_2, section1_3} of the bound electron--propagator
\bea
S_F (\mbox{\boldmath $r$}_1,\mbox{\boldmath $r$}_2,\omega) &=&
S_F^0 (\mbox{\boldmath $r$}_1-\mbox{\boldmath $r$}_2,\omega)
+\int d^3 \mbox{\boldmath $r$}\; 
S_F^0 (\mbox{\boldmath $r$}_1-\mbox{\boldmath $r$},\omega)
\gamma_0 V(\mbox{\boldmath $r$}) 
S_F^0 (\mbox{\boldmath $r$}-\mbox{\boldmath $r$}_2,\omega) \nonumber\\
&&\hspace{-3.0cm}+ \int d^3 \mbox{\boldmath $r$} \int d^3 \mbox{\boldmath $r'$}\;
S_F^0 (\mbox{\boldmath $r$}_1-\mbox{\boldmath $r$},\omega) \gamma_0 V(\mbox{\boldmath $r$})
S_F (\mbox{\boldmath $r$},\mbox{\boldmath $r'$},\omega) 
\gamma_0 V(\mbox{\boldmath $r'$}) 
S_F^0 (\mbox{\boldmath $r'$}-\mbox{\boldmath $r$}_2,\omega)
\label{section1_5}
\eea
that is represented graphically in Fig. \ref{fig2}. 
After this it is possible to apply the standard renormalization prescription
for the free QED. We use the BPHZ--renormalization--method 
\cite{section1_4,section1_5}. For any given divergent loop--integral 
$\hat {F}_{\Gamma}$ corresponding to a Feynman diagram $\Gamma$ this approach 
allows for a systematic isolation of the divergent parts and to derive the 
finite contribution $\hat {F}_{\Gamma}^{'}$ of the loop--integral under 
consideration. This procedure is formally expressed in terms of 
Bogoljubov's $\cal R$--operation acting onto the integrand $\hat {I}_{\Gamma}$ 
of the divergent loop--integral $\hat {F}_{\Gamma}$. This operation determines the 
integrand $\hat {R}_{\Gamma}$, which yields the finite integral $\hat {F}_{\Gamma}^{'}$:
\bea
\hat {F}_{\Gamma} = \int d^4 q_1 \; ... \; d^4 q_n \; \hat {I}_{\Gamma}
\quad \longrightarrow  \quad
\hat {F}_{\Gamma}^{'} = \int d^4 q_1 \; ...\; d^4 q_n \; \hat {R}_{\Gamma}\;.
\label{section1_15}
\eea
Depending on the superficial degree of divergence $\omega (\Gamma)$ of the diagram 
$\Gamma$ the action of Bogoljubov's $\cal R$--operation is defined recursively via 
\bea
\hat {R}_{\Gamma} = \left \{
\begin{array}[c]{l}
(1-t^{\Gamma}) \hat {\overline R}_{\Gamma} \quad \mbox{if} \quad 
\omega (\Gamma) \ge 0 \\
\displaystyle
\quad \quad \quad \; \,\hat {\overline R}_{\Gamma} \quad 
\mbox{if} \quad \omega (\Gamma) < 0 \\
\end{array} \right \}
\label{section1_20}
\eea
where Bogoliubov's recursion formula is given by
\bea
\hat {\overline R}_{\Gamma} = \hat {I}_{\Gamma} + \sum \limits_{\{\gamma_1,...,\gamma_k\}}
\hat {I}_{\Gamma/\{\gamma_1,...,\gamma_k\}} \prod \limits_{\tau=1}^{k}
(- t^{\gamma_{\tau}}) \hat {\overline R}_{\gamma_{\tau}} \quad .
\label{section1_25}
\eea
Here $\omega (\gamma) = 4 - E_B - \frac{3}{2} E_F$ defines the superficial  
degree of divergence for QED where $E_B$ and $E_F$ denote the number of external 
bosonic and fermionic legs, respectively, asigned to an arbitrary diagram or 
subdiagram $\gamma$. The sum in eq. (\ref{section1_25}) runs over all combinations of disjunct
superficially divergent subdiagrams $\gamma_i$ of the diagram 
$\Gamma$, and $\gamma_{\tau}$ denote the superficially divergent subdiagrams
of the subdiagrams $\gamma_i$. 
The mathematical expressions for the subdiagrams are obtained by means of the 
usual Feynman rules.
The symbol $\hat I_{\Gamma/{\gamma_1,...,\gamma_k}}$ stands for the integrand 
of the diagram $\Gamma$ where the subdiagrams ${\gamma_1,...,\gamma_k}$ are 
merged to a point.
The operator $t^{\gamma}$ denotes the Taylor expansion with respect to the 
independent external electron momenta $p_i$ and photon momenta $k_j$ up to 
order $\omega (\gamma)$ of the diagram $\gamma$. According to the 
renormalization group equation of QED the reference point of this Taylor expansion 
can be choosen arbitrary. However, a special choice of a renormalization scheme 
with fixed reference point implies renormalization conditions which are kept 
fixed. Here we employ on--shell--renormalization 
conditions and utilize the gauge invariance of the coupling between electron--positron 
field and the radiation field. This implies the following conditions for the 
renormalized electron self energy, the renormalized vertex function and for the 
second-- and fourth--rank vacuum polarization tensors of the photon, respectively, 
in perturbation theory of order $(\alpha)^n$ ($n=1,2,3,...$):
\bea
&&\hat {\Sigma}^{(n)\,{\rm ren}} (p) \bigg|_{\sla p = m} = 0 \;,\quad\quad
\hat {\Lambda}^{(n)\,{\rm ren}}_{\mu} (p_1,p_2) \bigg|_{\sla p_1=\sla p_2=m} 
= i e \gamma_{\mu} \;,\nonumber\\
\nonumber\\
&&\hat \Pi^{(n)\,{\rm ren}}_{\mu \nu} (k) \bigg|_{k=0} = 0 \;,\quad\quad
\hat \Pi^{(n)\,{\rm ren}}_{\mu \nu \rho \sigma} (k_1,k_2,k_3) 
\bigg|_{k_1=k_2=k_3=0} = 0 \;,
\label{section1_30}
\eea
where $m$ denotes the physical electron mass and $k_i=0$ means actually 
$k_i=(0,0,0,0)$. Therefore from the renormalization--conditions 
(\ref{section1_30}) one can define the 
Taylor--Operator $t^{\gamma}$ up to first order according to the prescription: \\
The operator $t^{\gamma}$ expands the integrand belonging to the diagram 
$\gamma$ 
\begin{enumerate}
\item[a)]
with respect to all independent $N_F$ external electron momenta $\sla p_i$ around $m$ 
\item[b)]
with respect to all independent $N_B$ external photon momenta $k_j$ around $0$.
\end{enumerate}
In our context here it will be sufficient to perform the Taylor expansion up to 
first order in external momenta. In mathematical terms we can write 
\bea
t^{\gamma} \hat {I}_{\gamma} (p_i,k_j) 
&=& \hat {I}_{\gamma} (p_i,k_j)\bigg|_{\sla p_i=m,k_j=0}\nonumber\\
\nonumber\\
&&\hspace{-3.0cm}+ \sum \limits_{i=1}^{N_F} 
\left( \frac {\partial}{\partial p_{i}^{\mu}} \hat {I}_{\gamma}
(p_i,k_j)\bigg|_{\sla p_i=m,k_j=0} \right) \left( p_{i}^{\mu} - m \frac{\gamma^{\mu}}{4} \right)
+ \sum \limits_{j=1}^{N_B} 
\left( \frac {\partial}{\partial k_{j}^{\mu}} \hat {I}_{\gamma}
(p_i,k_j)\bigg|_{\sla p_i=m,k_j=0} \right) k_j^{\mu} \nonumber\\
\nonumber\\
&&\hspace{-3.0cm}+ \,.\,.\,.\;\;.\nonumber\\
\label{section1_35}
\eea
\section{Renormalization of VPVP}

The contribution of the K\"all\'en--Sabry--diagrams with the free electron 
propagator are well--known \cite{intro7,intro8}. Accordingly we subtract these 
terms from the two--photon vacuum polarization diagram 
and concentrate on the higher--order contribution of VPVP which is 
graphically shown in the Fig. \ref{fig1}. As the major result in this chapter 
we derive the corresponding renormalized energy shift. 
\subsection{The outer renormalization}

According to the BPHZ--renormalization scheme as defined above the case where 
the Taylor--Operator $t^{\Gamma}$ acts on the whole diagram $\Gamma$ may be 
called outer renormalization. From Eq. (\ref{section1_20}) it becomes evident 
that outer renormalization is always required for a superficial degree of divergence 
$\omega (\Gamma) \ge 0$.
In the case of the VPVP diagram we have the following situation:\\
In the potential expansion of the energy shift in Fig. \ref{fig1} diagrams with at 
least four external Coulomb--photon legs occur. 
According to the Furry's theorem diagrams with an odd number of external 
photon or bosonic potential legs vanish. The superficial degree of 
divergence $\omega$ is zero for diagrams with four external bosonic legs 
and is negative for all diagrams of higher--order in $(Z\alpha)^n$ with $n\!>\!4$.
In view of Eq. (\ref{section1_20}) outer renormalization is required if and 
only if the number of outer bosonic legs is equal to four. 
These diagrams have to be considered with special care. \\
At this point we note that in the case of the one--loop vacuum polarization VP 
one must consider the diagram with four outer bosonic legs with special care 
as well (see for example \cite{section2_1}). Employing any regularization 
scheme in the one--loop case VP the corresponding counter term vanishes. 
Moreover, if the spherical expansion for the free and bound 
electron propagator is used no spurious non--gauge invariant term occurs 
in the 1--loop case VP \cite{section2_2}. For detailed discussions we refer to 
the literature (\cite{section2_1,section2_2,section2_3} and \cite{section2_4}). 
With this in mind one could expect that, after any regularization is performed, 
the counter terms required for the outer renormalization may vanish in the 
two--loop--case VPVP as well. In any case an explicit proof is desirable.\\
We turn to the investigation of the couterterms of the outer renormalization 
for the VPVP diagram. Performing the potential expansion of the energy shift 
shown in Fig. \ref{fig1} leads to six diagrams with $\omega=0$ (see
Fig. \ref{fig3}). To give an explicit example we consider graph $\Gamma_{\rm A}$ 
depicted in Fig. \ref{fig4}. This diagram includes the subdiagram 
$\gamma_{\rm A}$. It's contribution to the energy shift 
of an arbitrary bound electron state $\varphi_n$ reads 
\bea
\bigtriangleup E_{\rm n}^{\Gamma_A} 
&=& e^2 \int d^3 \mbox{\boldmath $r$} \;
\overline\varphi_{\rm n} (\mbox{\boldmath $r$}) 
\int \frac{d^3 \mbox{\boldmath $k$}_1}{(2 \pi)^3} 
\int \frac{d^3 \mbox{\boldmath $k$}_2}{(2 \pi)^3} 
\int \frac{d^3 \mbox{\boldmath $k$}_3}{(2 \pi)^3} \;
{\rm e}^{i (\mbox{\scriptsize\boldmath $k$}_1 + \mbox{\scriptsize\boldmath $k$}_2 + \mbox{\scriptsize\boldmath $k$}_3) 
\mbox{\scriptsize\boldmath $r$}} \nonumber\\
\nonumber\\
&&\hspace{-3.0cm}\;\times V (\mbox{\boldmath $k$}_1) \;V (\mbox{\boldmath $k$}_2)\;
V (\mbox{\boldmath $k$}_3)\;
\frac{1}{0^2 - (\mbox{\boldmath $k$}_1 + \mbox{\boldmath $k$}_2 + 
\mbox{\boldmath $k$}_3)^2 + i \epsilon} \;
\gamma_{0}\;\hat F_{\Gamma_{\rm A}} (\mbox{\boldmath $k$}_1,\mbox{\boldmath $k$}_2,
\mbox{\boldmath $k$}_3) \; \varphi_{\rm n} (\mbox{\boldmath $r$})\;,
\label{section2_5} 
\eea
where $V(\mbox{\boldmath $k$}_j)\;(j=1,2,3)$ is the Fourier--transform of 
the Coulomb potential of the nucleus:
\bea
V (\mbox{\boldmath $k$}_j) &=& \int d^3 \mbox{\boldmath $r$}\;
{\rm e}^{-i\mbox{\scriptsize\boldmath $k$}_j \mbox{\scriptsize\boldmath $r$}} \;
V (\mbox{\boldmath $r$})\;.
\label{section2_10}
\eea
The explicit mathematical expression for the Feynman diagram is given by 
(with $k_i = (0,\mbox{\boldmath $k$}_i)\;,i=1,2,3$)
\bea
&&\hat F_{\Gamma_{\rm A}} (\mbox{\boldmath $k$}_1,\mbox{\boldmath $k$}_2,
\mbox{\boldmath $k$}_3) 
= \int \frac {d^4 q_1}{(2 \pi)^4} \frac {d^4 q_2}{(2 \pi)^4}\;
I_{\Gamma_{\rm A}} (q_1,q_2,\mbox{\boldmath $k$}_1,\mbox{\boldmath $k$}_2,
\mbox{\boldmath $k$}_3) \;,\nonumber\\
\nonumber\\
&&I_{\Gamma_{\rm A}} (q_1,q_2,\mbox{\boldmath $k$}_1,\mbox{\boldmath $k$}_2,\mbox{\boldmath $k$}_3) 
= \nonumber\\
\nonumber\\
&&\times \, i\, {\rm Tr} \Bigg[
\gamma_{0} \frac{1}{\sla q_1+\sla k_1+\sla k_2 +
\sla k_3-m+i\epsilon} \gamma_0 \frac{1}{\sla q_1+\sla k_2 +\sla k_3
-m+i\epsilon} \gamma_0 \frac{1}{\sla q_1+\sla k_3-m+i\epsilon} \nonumber\\
\nonumber\\
&&\times\left(-i e^2 \frac {1}{q_2^2+i\epsilon}
\gamma_{\alpha}\frac{1}{\sla q_1-\sla q_2+\sla k_3-m+i\epsilon}
\gamma^{\alpha}\right)
\frac{1}{\sla q_1+\sla k_3-m+i\epsilon} \gamma_0
\frac{1}{\sla q_1-m+i\epsilon}
\Bigg]\;.\nonumber\\
\nonumber\\
\label{section2_15}
\eea
In round brackets the mathematical expression of integrand 
refering to the subdiagram $\gamma_{\rm A}$ (one--loop--self energy) 
has been inserted. Applying the $\cal R$--operation Eq. (\ref{section1_20}) 
and (\ref{section1_25}) yields 
\bea
\hat R_{\Gamma_{\rm A}} &=& \hat R_{\Gamma_{\rm A}}^{\;\rm inner} 
+ \hat R_{\Gamma_{\rm A}}^{\;\rm outer}
\eea
with
\bea
&&\hat R_{\Gamma_{\rm A}}^{\;\rm inner} =
\hat I_{\Gamma_{\rm A}} - \hat I_{\Gamma_{\rm A}/\gamma_{\rm A}}
\left(t^{\gamma_{\rm A}} \hat I_{\gamma_{\rm A}}\right)\;,\nonumber\\
\nonumber\\
&&\hat R_{\Gamma_{\rm A}}^{\;\rm outer} = 
-t^{\Gamma_{\rm A}} \hat I_{\Gamma_{\rm A}} + t^{\Gamma_{\rm A}} 
\left(\hat I_{\Gamma_{\rm A}/\gamma_{\rm A}}
\left(t^{\gamma_{\rm A}} \hat I_{\gamma_{\rm A}}\right)\right)\;.
\label{section2_35}
\eea
The Taylor--Operator $t^{\Gamma_{\rm A}}$ acts up to zeroth order while  
the other one $t^{\gamma_{\rm A}}$ acts up to first order since 
$\omega (\Gamma_{\rm A})=0$ and $\omega (\gamma_{\rm A})=1$.
Accordingly, the terms for the outer renormalization are
\bea
\int \frac {d^4 q_1}{(2 \pi)^4} \frac {d^4 q_2}{(2 \pi)^4}\;t^{\Gamma_{\rm A}}\left[
\hat I_{\Gamma_{\rm A}} (q_1,q_2,\mbox{\boldmath $k$}_1,\mbox{\boldmath $k$}_2,\mbox{\boldmath $k$}_3)\right]
= \int \frac {d^4 q_1}{(2 \pi)^4} \frac {d^4 q_2}{(2 \pi)^4}\;
\hat I_{\Gamma_{\rm A}} (q_1,q_2,\mbox{\boldmath $0$},\mbox{\boldmath $0$},\mbox{\boldmath $0$})
\label{section2_40}
\eea
and
\bea
\int \frac {d^4 q_1}{(2 \pi)^4} \frac {d^4 q_2}{(2 \pi)^4}\;
t^{\Gamma_{\rm A}} 
\left[\hat I_{\Gamma_{\rm A}/\gamma_{\rm A}} (q_1,\mbox{\boldmath $k$}_1,\mbox{\boldmath $k$}_2,\mbox{\boldmath $k$}_3)
\left(t^{\gamma_{\rm A}}
\hat I_{\gamma_{\rm A}} \left(q_1,q_2,\mbox{\boldmath $k$}_1,\mbox{\boldmath $k$}_2,\mbox{\boldmath $k$}_3\right)\right)\right]\nonumber\\
\nonumber\\
&&\hspace{-12.5cm}= \Sigma^{(1)} \int \frac {d^4 q_1}{(2 \pi)^4}\;
\hat I_{\Gamma_{\rm A}} (q_1,\mbox{\boldmath $0$},\mbox{\boldmath $0$},\mbox{\boldmath $0$}) \;
+ \;\Sigma^{(1)'} \int \frac {d^4 q_1}{(2 \pi)^4}\;
\underline{\left(\sla q_1 - m \right)
\hat I_{\Gamma_{\rm A}/\gamma_{\rm A}}} (q_1,q_2,\mbox{\boldmath $0$},\mbox{\boldmath $0$},\mbox{\boldmath $0$})\;,\nonumber\\
\label{section2_45}
\eea
respectively. The denotation $\underline{\left(\sla q_1 - m \right)
\hat I_{\Gamma_{\rm A}/\gamma_{\rm A}}} $ implies insertion of the 
Dirac--structure $(\sla q_1 - m)$ at the same place
in the diagram $\Gamma_{\rm A}/\gamma_{\rm A}$ where there was the 
subdiagram $\gamma_{\rm A}$ before.
In Eq. (\ref{section2_45}) we used the usual definition of the 
one--loop--counter terms $\Sigma^{(1)}$ and $\Sigma^{(1)'}$ 
of free QED:
\bea
\Sigma^{(1)} = - i e^2 \int \frac{d^4 q_2}{(2 \pi)^4}
\gamma_{\alpha} \frac{1}{\sla q_1-\sla q_2-m+i\epsilon} 
\gamma^{\alpha} \frac{1}{q_2^2+i\epsilon}\bigg|
_{\sla q_1=m} 
\label{section2_47}
\eea
and
\bea
\gamma_{\rho} \Sigma^{(1)'} = - i e^2 \frac {\partial}{\partial q_1^{\rho}}
\int \frac{d^4 q_2}{(2 \pi)^4}
\gamma_{\alpha} \frac{1}{\sla q_1-\sla q_2-m+i\epsilon} 
\gamma^{\alpha} \frac{1}{q_2^2+i\epsilon}\bigg|
_{\sla q_1=m}\;.
\label{section2_48}
\eea
Similar steps have to be performed in all other diagrams of Fig. \ref{fig3}. 
With the aid of the one--loop Ward--identity
\bea
\frac{\partial\hat\Sigma^{(1){\rm ren}}(q_1)}{\partial q_1^0} = 
\hat \Lambda_0^{(1){\rm ren}} (q_1,q_1)\;,\quad
\frac{\partial \hat\Sigma^{(1)}(q_1)}{\partial q_1^0} = 
\hat \Lambda_0^{(1)} (q_1,q_1)\;,\quad
\Sigma^{(1)'} = - \Lambda^{(1)}
\label{section2_50}
\eea
together with the identity 
\bea
\underline{\left(\sla q_1-m \right) \hat I_{\Gamma_{\rm A}/\gamma_{\rm A}}} = 
\underline{\gamma_0 \hat I_{\Gamma_{\rm B}/\gamma_{\rm B}}}
\label{section2_55}
\eea
and the relations between renormalized and unrenormalized vertex and 
self--energy operator 
\bea
\hat \Lambda^{(1)}_0 (\sla q_1 - \sla q_2, \sla q_1 - \sla q_2) &=&
\gamma_0 \Lambda^{(1)} + \hat \Lambda_{0}^{(1) {\rm ren}} 
(\sla q_1 - \sla q_2, \sla q_1 - \sla q_2)\;, \nonumber\\
\nonumber\\
\hat \Sigma^{(1)} (\sla q_1) &=& \Sigma^{(1)} 
+ \left(\sla q_1 - m \right) \Sigma^{(1)'}
+ \hat \Sigma^{(1) {\rm ren}} (\sla q_1)\;,
\label{section2_60}
\eea
all counter terms refering to the outer renormalization can be collected. 
One ends up with the following expression for the sum of all 
counter terms generated by the operator $\hat R^{\;\rm outer}_{\Gamma}$ 
corresponding to the six diagrams of Fig. \ref{fig3}:
\bea
&&\int\frac{d^4 q_1}{(2\pi)^4} \int\frac{d^4 q_2}{(2\pi)^4} 
\left(2 \hat R_{\Gamma_{\rm A}}^{\;\rm outer} 
+ 2 \hat R_{\Gamma_{\rm B}}^{\;\rm outer} 
+ 2 \hat R_{\Gamma_{\rm C}}^{\;\rm outer} 
+ \hat R_{\Gamma_{\rm D}}^{\;\rm outer} 
+ \hat R_{\Gamma_{\rm E}}^{\;\rm outer} 
+ \hat R_{\Gamma_{\rm F}}^{\;\rm outer} \right)\nonumber\\
\nonumber\\
&=&{\rm Tr}\;\int \frac{d^4 q_1}{(2 \pi)^4}  
\frac{\partial}{\partial q_{1}^{0}}
\Bigg[\frac{1}{\sla q_1 - m +i\epsilon} 
\hat\Lambda_{0}^{(1){\rm ren}} (\sla q_1, \sla q_1) 
\frac{1}{\sla q_1 - m +i\epsilon} \gamma_0
\frac{1}{\sla q_1 - m +i\epsilon} \gamma_0\nonumber\\
\nonumber\\
&&+\frac{1}{\sla q_1 - m +i\epsilon} 
\hat\Sigma^{(1){\rm ren}} (\sla q_1) 
\frac{1}{\sla q_1 - m +i\epsilon} \gamma_0
\frac{1}{\sla q_1 - m +i\epsilon} \gamma_0
\frac{1}{\sla q_1 - m +i\epsilon} \gamma_0\Bigg]\;. \nonumber\\
\label{section2_70}
\eea
The last integral is understood as being regularized.
The renormalized self energy as well as the renormalized 
vertex correction do not posess any singularities. Therefore the
total integrand has simple poles only at 
$q_1^0=\pm\sqrt{\mbox{\boldmath $q$}_1^2+(m-i\epsilon)^2}$. Accordingly, 
a Wick--rotation $q_1^0 \rightarrow i q_1^0$ to Euclidean space can be 
carried out and all surface terms vanishes after regularization. 
This implies that all counter terms generated by $\hat R^{\;\rm outer}_{\Gamma}$ 
cancel each other. Therefore all singularities associated with the outer 
renormalization must cancel in the unrenormalized expression of VPVP.

\subsection{The inner renormalization}

According to the previous subsection only an inner renormalization associated with 
the action of $\hat R^{\;\rm inner}_{\Gamma}$ is required. For this 
pupose we perform the potential expansion of the 
bound--electron propagator until three outer bosonic potential legs. 
An appropriate potential expansion of the energy correction of higher--order 
in eq. (\ref{fig1}) is given in Fig. \ref{fig5}.
For the last two diagrams C2 and C3 an inner renormalization
is not adequate. For the other six diagrams A1, A2, A3, B1, B2 and C1 
a simple inner renormalization is necessary. There are the inner divergencies 
of the one--loop--self energy as well as the one--loop--vertex correction.
The renormalization of this one--loop--expressions has to be carried out 
using eq. (\ref{section2_60}). Accordingly, one should perform the 
replacements of Fig. \ref{fig6} in order to receive the full renormalized expression 
for VPVP. If one uses the Ward--identity Eq. (\ref{section2_50}), 
Furry's theorem and the potential expansion shown in Fig. \ref{fig2}, it is 
not difficult to realize that all the terms which are proportional to $\Sigma^{(1)'}$ are 
cancelled against the terms which are proportional to $\Lambda^{(1)}$.
Especially, it holds the following relations for these counter terms:
\bea
2\; \Lambda^{(1)}\;{\rm CT} ({\rm B 1}) + 2\; \Sigma^{(1)'}\;{\rm CT} ({\rm A 1}) &=& 0 \;,\nonumber\\
\nonumber\\
\Lambda^{(1)}\;{\rm CT} ({\rm B 2}) + \Lambda^{(1)}\;{\rm CT} ({\rm C 1}) + 
2\;\Sigma^{(1)'}\;{\rm CT} ({\rm A 2}) + \Sigma^{(1)'}\;{\rm CT} ({\rm A 3}) 
&=& 0 \;.
\label{section2_75}
\eea
Finally, if one collects all terms proportional to $\Sigma^{(1)}$  and 
retransforms the potential expansion one ends up with the renormalized expression 
for the total energy shift of the VPVP, see Fig. \ref{fig7}.

\section{The renormalized expression of VPVP}

Now we proceed to write down the mathematical expression of the 
renormalized vacuum polarization correction. The energy shift 
in higher--order reads:
\bea
\bigtriangleup E_n^{\rm VPVP\;ren\;h.o.} &=& 
\int d^3 \mbox{\boldmath $r$}_1 \;\overline{\varphi}_n (\mbox{\boldmath $r$}_1) \;
\gamma_0\;\varphi_n (\mbox{\boldmath $r$}_1)\nonumber\\
\nonumber\\
&&\hspace{-3.0cm}\times\Bigg[\hat{U}^{\rm VPVP}(\mbox{\boldmath $r$}_1)
-2 \hat{U}^{\rm A0}(\mbox{\boldmath $r$}_1)
-\hat{U}^{\rm B0}(\mbox{\boldmath $r$}_1)
-\Sigma^{(1)} \left(\hat{U}^{\rm CT1}(\mbox{\boldmath $r$}_1)
-2\hat{U}^{\rm CT2}(\mbox{\boldmath $r$}_1)\right)\Bigg]\;.\nonumber\\
\nonumber\\
\label{section3_5}
\eea
It is depicted in Fig. \ref{fig7}.
The potentials are given by (note that $\Sigma^{(1)}$ is proportional to $e^2$)
\bea
\hat U^{\rm VPVP} ({\mbox{\boldmath $r$}_1}) &=& {\rm e}^4 \,
 \int d^3 {\mbox{\boldmath $r$}_2} \int d^3 {\mbox{\boldmath $r$}_3} \int d^3 {\mbox{\boldmath $r$}_4} 
\int\limits_{-\infty}^{\infty} \frac{d E_1}{2\pi} 
\int\limits_{-\infty}^{\infty} \frac{d E_2}{2\pi} \nonumber\\
\nonumber\\
&&\times D^{00} (E_1, {\mbox{\boldmath $r$}_1}-{\mbox{\boldmath $r$}_2}) D^{\alpha \beta} 
(E_2, {\mbox{\boldmath $r$}_3}-{\mbox{\boldmath $r$}_4}) \nonumber\\
\nonumber\\
&&\times{\rm Tr} \left[ \gamma_0 S_F (E_1, {\mbox{\boldmath $r$}_2},{\mbox{\boldmath $r$}_3})
\gamma_{\alpha} S_F(E_1-E_2,{\mbox{\boldmath $r$}_3},{\mbox{\boldmath $r$}_4}) 
\gamma_{\beta} S_F (E_1,{\mbox{\boldmath $r$}_4},{\mbox{\boldmath $r$}_2})\right]\;.\nonumber\\
\label{section3_10}
\eea
\bea
\hat U^{\rm A0} ({\mbox{\boldmath $r$}_1}) &=& 
e^4 \int d^3 {\mbox{\boldmath $r$}_2}\; d^3 {\mbox{\boldmath $r$}_3}\; d^3 {\mbox{\boldmath $r$}_4} \;d^3 {\mbox{\boldmath $r$}_5} 
\int \limits_{- \infty}^{\infty} \frac{d E_1}{2 \pi}
\int \limits_{- \infty}^{\infty} \frac{d E_2}{2 \pi}\nonumber\\
\nonumber\\
&&\times D^{00} ({\mbox{\boldmath $r$}_1} - {\mbox{\boldmath $r$}_2}, 0)
D^{\alpha \beta} ({\mbox{\boldmath $r$}_3} - {\mbox{\boldmath $r$}_4}, E_2) V ({\mbox{\boldmath $r$}_5}) \nonumber\\
\nonumber\\
&&\times \mbox{Tr}\, \bigg [ S_F^0
({\mbox{\boldmath $r$}_2}-{\mbox{\boldmath $r$}_3}, E_1) \gamma_{\alpha}
S_F^0 ({\mbox{\boldmath $r$}_3}-{\mbox{\boldmath $r$}_4}, E_1-E_2) \gamma_{\beta}
S_F^0 ({\mbox{\boldmath $r$}_4}-{\mbox{\boldmath $r$}_5}, E_1)\nonumber\\
\nonumber\\
&&\times S_F^0 ({\mbox{\boldmath $r$}_5}-{\mbox{\boldmath $r$}_2}, E_1) \bigg]\;,
\label{section3_15}
\eea
\bea
\hat U^{\rm B0} ({\mbox{\boldmath $r$}_1}) &=&
e^4 \int d^3 {\mbox{\boldmath $r$}_2} \;d^3 {\mbox{\boldmath $r$}_3}\; d^3 {\mbox{\boldmath $r$}_4} \;d^3 {\mbox{\boldmath $r$}_5} 
\int \limits_{- \infty}^{\infty} \frac{d E_1}{2 \pi}
\int \limits_{- \infty}^{\infty} \frac{d E_2}{2 \pi}\nonumber\\
\nonumber\\
&&\times D^{00} ({\mbox{\boldmath $r$}_1} - {\mbox{\boldmath $r$}_2}, 0)
D^{\alpha \beta} ({\mbox{\boldmath $r$}_3} - {\mbox{\boldmath $r$}_4}, E_2) V ({\mbox{\boldmath $r$}_5}) \nonumber\\
\nonumber\\
&&\times \mbox{Tr}\, \bigg[ S_F^0
({\mbox{\boldmath $r$}_2}-{\mbox{\boldmath $r$}_3}, E_1) \gamma_{\alpha}
S_F^0 ({\mbox{\boldmath $r$}_3}-{\mbox{\boldmath $r$}_5}, E_1 - E_2)
S_F^0 ({\mbox{\boldmath $r$}_5}-{\mbox{\boldmath $r$}_4}, E_1 - E_2)
\gamma_{\beta} \nonumber\\
\nonumber\\
&&\times S_F^0 ({\mbox{\boldmath $r$}_4}-{\mbox{\boldmath $r$}_2}, E_1) \bigg]\;,
\label{section3_20}
\eea
\bea
\hat U^{\rm CT1} ({\mbox{\boldmath $r$}_1}) &=&
e^2 \int d^3 {\mbox{\boldmath $r$}_2}\; d^3 {\mbox{\boldmath $r$}_3} 
\int\limits_{- \infty}^{\infty} \frac{d E}{2 \pi}
D^{0 0} ({\mbox{\boldmath $r$}_1} - {\mbox{\boldmath $r$}_2}, 0) 
\mbox{Tr}\, \left [ S_F ({\mbox{\boldmath $r$}_2},{\mbox{\boldmath $r$}_3}, E) \gamma_0 
S_F ({\mbox{\boldmath $r$}_3},{\mbox{\boldmath $r$}_2}, E)\right]\;,
\nonumber
\eea
\vspace{-0.5cm}
\bea
\label{section3_25}
\eea
\bea
\hat {U}^{\rm CT2} ({\mbox{\boldmath $r$}_1}) &=&
e^2 \int d^3 {\mbox{\boldmath $r$}_2}\; d^3 {\mbox{\boldmath $r$}_3} \;d^3 {\mbox{\boldmath $r$}_4}  
\int \limits_{- \infty}^{\infty} \frac{d E}{2 \pi}
D^{0 0} ({\mbox{\boldmath $r$}_1} - {\mbox{\boldmath $r$}_2}, 0) V ({\mbox{\boldmath $r$}_3}) \nonumber\\
\nonumber\\
&&\times \mbox{Tr} \left [ \gamma_0 S_F^0
({\mbox{\boldmath $r$}_2}-{\mbox{\boldmath $r$}_3}, E) 
S_F^0 ({\mbox{\boldmath $r$}_3}-{\mbox{\boldmath $r$}_4}, E)
S_F^0 ({\mbox{\boldmath $r$}_4}-{\mbox{\boldmath $r$}_2}, E)\right]\;.
\label{section3_30}
\eea
In the expressions above $S_F$ denotes the bound--electron propagator and 
$S_F^0$ denotes the free--electron propagator, respectively, and 
$D^{\mu \nu}$ is the photon propagator (see appendix). 




\section{The fourth--rank vacuum polarization tensor}

In order to derive a renormalized expression for the self energy vacuum 
polarization S(VP)E it is useful to mention some properties of the 
fourth--rank vacuum polarization tensor. 
Especially we reexamine some features of this tensor which are related to the 
gauge invariance of the vacuum polarization diagram VP. 
The fourth--rank vacuum polarization tensor is defined as amplitude of the
light--light scattering which is shown in Fig. \ref{fig9}. The 
expression of the unrenormalized tensor reads:
\bea
\hat {\Pi}^{\rm unren}_{\mu,\nu,\sigma,\lambda} (k_1,k_2,k_3,m) 
&=& 2\bigg[ \hat{T}^{\rm unren}_{\mu,\nu,\lambda,\sigma}(k_1,k_2,k_3,m)  
+ \hat{T}^{\rm unren}_{\mu,\nu,\sigma,\lambda}(k_1,k_2,-k_1-k_2-2 k_3,m)\nonumber\\
\nonumber\\
&& +\;\hat{T}^{\rm unren}_{\mu,\lambda,\nu,\sigma}(k_1,k_3,k_2,m)
\bigg]\;,
\label{eq_tensor_5}
\eea
where the single three tensors $\hat{T}^{\rm unren}$ are corresponding to the 
expressions belonging to the diagrams depicted in Fig. \ref{fig9} and they 
are defined by
\bea
\hat{T}^{\rm unren}_{\mu,\nu,\lambda,\sigma}(k_1,k_2,k_3,m) 
&=& \int \frac{d^4 q}{(2 \pi)^4} {\rm Tr}\;
\bigg( \gamma_{\mu} \frac{1}{\sla q+\sla k_1-m+i\epsilon}\gamma_{\nu}
\frac{1}{\sla q+\sla k_1+\sla k_2-m+i\epsilon}\gamma_{\lambda} \nonumber\\
\nonumber\\
&&\hspace{-3.0cm}\times\frac{1}{\sla q+\sla k_1+\sla k_2+\sla k_3-m+i\epsilon}\gamma_{\sigma}
\frac{1}{\sla q-m+i\epsilon}\bigg)\;.
\label{eq_tensor_10}
\eea
For our purposes it is necessary to consider the general case where the 
external momenta $k_i$ are not on mass shell ($k_i^2\neq 0$).
The general structure of the solution for this tensor is complicated and was 
derived at the first time by {\it Karplus, Neumann} in \cite{lit_2_5}.
For our intension here it is only important that the single tensors 
$\hat T^{\rm unren}_{\mu,\nu,\sigma,\lambda}$ 
are logarithmically divergent but their sum $\hat \Pi^{\rm unren}_{\mu,\nu,\sigma,\lambda}$ 
is convergent.
This surprising behaviour of $\hat \Pi^{\rm unren}_{\mu,\nu,\sigma,\lambda}$ has been 
proven a long time ago \cite{lit_2_10,lit_2_15}, see also \cite{section2_3}. 
Therefore a renormalization of $\hat \Pi^{\rm unren}_{\mu,\nu,\sigma,\lambda}$ 
seems to be not necesarry, which, however, is not true. The vacuum polarization 
tensor $\hat \Pi^{\rm unren}_{\mu,\nu,\sigma,\lambda}$ is not gauge invariant 
and therefore a renormalization is necesarry. Following the BPHZ--renormalization 
method described in section 2, we obtain the renormalized expression 
\bea
\hat {\Pi}^{\rm ren}_{\mu,\nu,\sigma,\lambda} (k_1,k_2,k_3,m) 
&=& \hat {\Pi}^{\rm unren}_{\mu,\nu,\sigma,\lambda} (k_1,k_2,k_3,m) 
- \hat {\Pi}^{\rm unren}_{\mu,\nu,\sigma,\lambda} (0,0,0,m) \;.
\label{eq_tensor_15}
\eea
Gauge invariance implies (see for example \cite{lit_2_15,section2_3,
section1_5,lit_2_26} 
and references therein)
\bea
k_1^{\mu}\,\hat {\Pi}^{\rm ren}_{\mu,\nu,\sigma,\lambda} (k_1,k_2,k_3,m) 
= 0 \;,\nonumber\\
k_2^{\nu}\,\hat {\Pi}^{\rm ren}_{\mu,\nu,\sigma,\lambda} (k_1,k_2,k_3,m) 
= 0 \;,\nonumber\\
k_3^{\sigma}\,\hat {\Pi}^{\rm ren}_{\mu,\nu,\sigma,\lambda} (k_1,k_2,k_3,m) 
= 0 \;.
\label{eq_tensor_20}
\eea
In other words the counter term 
$\hat {\Pi}^{\rm unren}_{\mu,\nu,\sigma,\lambda} (0,0,0,m)$ ensure the 
gauge invariance of the fourth--rank vacuum polarization tensor.
We note that the regularized tensor 
\bea
\hspace{-1.0cm}\hat {\Pi}^{\rm reg}_{\mu,\nu,\sigma,\lambda} (k_1,k_2,k_3,m,M) 
&=& \hat {\Pi}^{\rm unren}_{\mu,\nu,\sigma,\lambda} (k_1,k_2,k_3,m) 
- \hat {\Pi}^{\rm unren}_{\mu,\nu,\sigma,\lambda} (k_1,k_2,k_3,M)\;,
\label{eq_tensor_25}
\eea
is also gauge invariant ($M$ denotes a large electron mass of the 
Pauli--Villars--regula\-rization). A renormalization of the regularized tensor 
in Eq. (\ref{eq_tensor_25}) is not necesarry 
because the counter term $\hat {\Pi}^{\rm unren}_{\mu,\nu,\sigma,\lambda} (0,0,0,m)$ 
would vanish if one performs such a regularization. 
Sometimes the subtracted term $\hat {\Pi}^{\rm unren}_{\mu,\nu,\sigma,\lambda} (k_1,k_2,k_3,M)$ 
in Eq. (\ref{eq_tensor_25}),
which ensures the gauge invariance like the counter term in Eq. 
(\ref{eq_tensor_15}),
is called spurious term. Similarly we can consider the counter term 
$\hat {\Pi}^{\rm unren}_{\mu,\nu,\sigma,\lambda} (0,0,0,m)$ in 
Eq. (\ref{eq_tensor_15}) 
as a spurious term.\\
There are regularization methods, in which the counter term 
$\hat {\Pi}^{\rm unren}_{\mu,\nu,\sigma,\lambda} (0,0,0,m)$ in 
Eq. (\ref{eq_tensor_15}) 
or the subtracted term $\hat {\Pi}^{\rm unren}_{\mu,\nu,\sigma,\lambda} (k_1,k_2,k_3,M)$ 
in Eq. (\ref{eq_tensor_25})
will vanish \cite{section2_2,section2_4}. 
It means that these terms are not necesarry in such cases in order to ensure 
gauge invariance. Because of this it is important to note that both 
approaches alone, Eq. (\ref{eq_tensor_15}) as well as Eq. (\ref{eq_tensor_25}), 
ensure the gauge invariance of the fourth--rank vacuum polarization tensor. 
Therefore is does not play any role which term will vanish. 
However, because of the simplier mathematical structure of the spurious term 
of Eq. (\ref{eq_tensor_15}) in comparison with the spurious term 
of Eq. (\ref{eq_tensor_25}) 
it turns out to be much easier to use the approach of Eq. 
(\ref{eq_tensor_15}) instead of Eq. (\ref{eq_tensor_25}).  
As we will see 
in the next section the spherical expansion of electron propagator is such 
a special condition where the counter term in Eq. (\ref{eq_tensor_15}) 
will vanish.

\section{Renormalization of the S(VP)E diagram and disappearance 
         of the spurious term}

In order to isolate the ultraviolet divergencies of the S(VP)E diagram we 
first have to perform a potential expansion of the bound electron--propagator 
according to Eq. (\ref{section1_5}) as we have seen already in the case of the VPVP. 
For the diagram of the S(VP)E such an appropriate potential expansion is shown in 
Fig. \ref {fig11}. The diagrams 
A3, B2, C2 and D are convergent. The sum of diagrams A2, B1 and C1 are nothing 
else but the fourth--rank vacuum polarization tensor. Therefore they are separately 
divergent but their sum is convergent but not gauge invariant as we have discussed 
in the previous section. It should be mention here that only for hydrogen 
an evaluation of the diagrams A2, B1 and C1 was performed (see \cite{lit_3_20} 
and references therein).
At present, for all other elements of the periodic system their contribution  
is unknown. Diagram A1 is the so--called Uehling contribution of the S(VP)E.
This Uehling contribution was evaluated for low-- and high--$Z$ atomic systems 
few years ago \cite{lit_3_25} and gave the leading contribution 
at least for high--Z--systems like uranium and lead. \\
Now we proceed to prove the disappearance of counter terms of 
diagrams A2, B1 and C1 if one uses the spherical expansion of the 
free--electron propagator (see appendix). 
Their corresponding energy shift is given by:
\bea
E_n^{\rm (A2)+(B1)+(C1)} &=& e^4 \int\limits_{-\infty}^{\infty}\frac{d E_1}{2\pi}
\int\frac{d^3 {\mbox{\boldmath $k$}_1}}{(2\pi)^3} 
\int\frac{d^3 {\mbox{\boldmath $k$}_3}}{(2\pi)^3}\; V({\mbox{\boldmath $k$}_3})
\int\frac{d^3 {\mbox{\boldmath $k$}_4}}{(2\pi)^3}\; V({\mbox{\boldmath $k$}_4})\nonumber\\
\nonumber\\
&&\hspace{-3.5cm}\times\frac{1}{E_1^2-{\mbox{\boldmath $k$}_1}^2+i\epsilon}\;
\frac{1}{E_1^2-({\mbox{\boldmath $k$}_1}-{\mbox{\boldmath $k$}_3}-{\mbox{\boldmath $k$}_4})^2+i\epsilon}\;
\frac{1}{E_n-E_1-E_m (1-i\epsilon)}\nonumber\\
\nonumber\\
&&\hspace{-3.5cm}\times\sum\limits_m\int d^3 \mbox{\boldmath $r$}_1 \;
\varphi_n ({\mbox{\boldmath $r$}_1})\alpha^{\mu} 
{e}^{i{\mbox{\scriptsize\boldmath $k$}_1}
{\mbox{\scriptsize\boldmath $r$}_1}}\varphi_m ({\mbox{\boldmath $r$}_1}) \;
\int d^3 \mbox{\boldmath $r$}_2 \;\varphi_m ({\mbox{\boldmath $r$}_2})\alpha^{\nu} 
{e}^{-i{(\mbox{\scriptsize\boldmath $k$}_1+\mbox{\scriptsize\boldmath $k$}_3+
\mbox{\scriptsize\boldmath $k$}_4)}
{\mbox{\scriptsize\boldmath $r$}_2}}\varphi_n ({\mbox{\boldmath $r$}_2}) \nonumber\\
\nonumber\\
&&\hspace{-3.5cm}\times\hat{\Pi}_{\mu,\nu,0,0} (E_1,\mbox{\boldmath $k$}_1;E_1,
\mbox{\boldmath $k$}_1+\mbox{\boldmath $k$}_3+\mbox{\boldmath $k$}_4;0,
\mbox{\boldmath $k$}_3;0,\mbox{\boldmath $k$}_4;m)\;.
\label{eq_section1_10}
\eea
Using the free--electron propagator
\bea
S_F^0 (E,\mbox{\boldmath $r$}_1-\mbox{\boldmath $r$}_2) &=& \int\frac{d^3 {\mbox{\boldmath $k$}}}{(2\pi)^3}
\sum\limits_{i=1}^{4} \frac{\varphi_i (\mbox{\boldmath $k$}){e}^{i{\mbox {\scriptsize\boldmath $k\,r$}_1}}
\varphi_i^{\dagger} ({\mbox {\boldmath $k$}}){e}^{-i{\mbox {\scriptsize\boldmath $k\,r$}_2}}}{E-E_{k}(1-i\epsilon)}\;,
\label{eq_section1_15}
\eea
where the functions $\varphi_i ({\mbox {\boldmath $k$}})$ are the solutions 
of the free Dirac equation in momentum space:
\bea
\left(\sla k-m\right) \varphi_i ({\mbox {\boldmath $k$}}) &=& 0 \;,\quad
\varphi^{\dagger}_{i_1} ({\mbox{\boldmath $k$}}) \varphi_{i_2} ({\mbox{\boldmath $k$}}) 
= \delta_{i_1,i_2}\;,
\label{eq_section1_25}
\eea
(here the indices $i$ indicate the four linear independent solutions 
of the free Dirac equation) one may derive the relation
\bea
\frac{1}{E_2 \gamma_0-{\mbox{\boldmath $k$}_2}{\mbox{\boldmath $\gamma$}}-m+i\epsilon} 
&=& \sum\limits_{i=1}^{4} \frac{\varphi_{i} ({\mbox{\boldmath $k$}_2}) 
\varphi^{\dagger}_{i} ({\mbox{\boldmath $k$}_2})}
{E_2-E_{k_2,i}(1-i\epsilon)}\;.
\label{eq_section1_20}
\eea
The last three equations lead immediately to the following expression 
for the counter term of the fourth--rank vacuum polarization tensor in 
Eq. (\ref{eq_section1_10}):
\bea
\hspace{-1.0cm}\hat{\Pi}_{\mu,\nu,0,0} (0,0,0,0,m) &=&
\frac{1}{2}\int\limits_{-\infty}^{\infty}\frac{d E_2}{2\pi}\int d^3 ({\mbox{\boldmath $r$}_1-\mbox{\boldmath $r$}_2})\nonumber\\
\nonumber\\
&&\times{\rm Tr} \bigg[
\frac{\partial^2}{\partial E_2^2}\left(\alpha_{\mu}
S_F^0 (E_2,{\mbox{\boldmath $r$}_1-\mbox{\boldmath $r$}_2}) \alpha_{\nu} 
S_F^0 (E_2,{\mbox{\boldmath $r$}_2-\mbox{\boldmath $r$}_1})\right)\bigg]\;.
\label{eq_section1_30}
\eea
Using the identity ($\hat H_0^{\rm Dirac}$ is the free Dirac operator without 
Coulomb potential)
\bea
\alpha_j &=& \left(\hat H_0^{\rm Dirac}-E\right) r_j 
- r_j \left(\hat H_0^{\rm Dirac}-E\right) \;,
\quad j=1,2,3
\label{eq_section1_35}
\eea
and the relation 
\bea
\left(\hat H_0^{\rm Dirac}-E\right) S_F^0 (E,{\mbox{\boldmath $r$}_1-\mbox{\boldmath $r$}_2}) &=& \delta ({\mbox{\boldmath $r$}_1-\mbox{\boldmath $r$}_2})\;,
\label{eq_section1_40}
\eea
and their conjugate version, respectively, it is not difficult to see that 
the spatial components in Eq. (\ref{eq_section1_30}) do not contribute.
Finally, with $\alpha_0={\bf 1}\hspace{-4pt}1$ we get 
\bea
\hat{\Pi}_{\mu=0,\nu=0,0,0} (0,0,0,0,m) &=&
\frac{1}{6}\int\limits_{-\infty}^{\infty}\frac{d E_2}{2\pi}\;
{\rm Tr} \bigg[\frac{\partial^3}{\partial E_2^3}
\left(S_F^0 (E_2,{\mbox{\boldmath $r$}_1-\mbox{\boldmath $r$}_1})\right)\bigg]\;.
\label{eq_section1_45}
\eea
If one inserts the spherical expansion of the free--electron propagator 
(see appendix) one ends up with 
\bea
\hat{\Pi}_{\mu=0,\nu=0, 0, 0} (0,0,0,0,m) =
\frac{1}{6}\sum\limits_{\kappa}\frac{|\kappa|}{2\pi}
\int\limits_{-\infty}^{\infty}\frac{d E_2}{2\pi}
\frac{\partial^3}{\partial E_2^3}\left[
G^{11}_{0,\kappa}(E_2,r_1,r_1) + G^{22}_{0,\kappa}(E_2,r_1,r_1)\right]\;.
\nonumber
\eea
\vspace{-0.8cm}
\bea
\label{eq_section1_50}
\eea
The same expression was derived and investigated already in \cite{section2_2} 
where it has been shown that this expression vanishes if the summation over $\kappa$ 
is restricted to a finite number of terms.
This can be seen by inserting the explicit expression of the spherical expansion
where one gets after a Wick--rotation the following expression:
\bea
\hat{\Pi}_{\mu=0,\nu=0, 0, 0} (0,0,0,0,m) &=& 
\frac{1}{6}\sum\limits_{\kappa}\frac{|\kappa|}{2\pi}
\int\limits_{-\infty}^{\infty} \frac{d E_2}{2\pi} \frac{\partial^3}{\partial E_2^3}\;
{\rm Re} \left(i \left[G^{11}_{0,\kappa}(i E_2) + G^{22}_{0,\kappa}(i E_2)\right]\right)\nonumber\\
\nonumber\\
&&\hspace{-5.0cm}= -\frac{1}{6}\sum\limits_{\kappa}\frac{|\kappa|}{2\pi}
\frac{\partial^2}{\partial E_2^2} \nonumber\\
\nonumber\\
&&\hspace{-5.0cm}\times\Bigg[(i E_2+1)\sqrt{1+E_2^2} \;j_{|\kappa+1/2|-1/2} (i\sqrt{1+E_2^2}\;r_1)\; 
h_{|\kappa+1/2|-1/2}^{(1)} (i\sqrt{1+E_2^2}\;r_1)\nonumber\\
\nonumber\\
&&\hspace{-5.0cm}+(iE_2-1)\sqrt{1+E_2^2}\; j_{|\kappa-1/2|-1/2} (i\sqrt{1+E_2^2}\;r_1) \;
h_{|\kappa-1/2|-1/2}^{(1)} (i\sqrt{1+E_2^2}\;r_1)\Bigg]_{E_2=-\infty}^{E_2=\infty}\;.
\nonumber\\
\nonumber\\
\label{eq_section1_60}
\eea
Here $j_{|\kappa\pm 1/2|-1/2} (z)$ and $h_{|\kappa\pm 1/2|-1/2}^{(1)} (z)$ 
are the spherical Bessel function and spherical Hankel function of first 
kind, respectively.
In the last line we used the Gauss law in Euclidean space.
It can readily be seen that the last expression vanishes if one performs both 
derivations. So it has been proven that the counter term is zero if one uses 
the spherical 
expansion of the electron propagator where one has to sum over $\kappa$ 
in the last step. This result simplifies essentially  
the evaluation of arbitrary diagrams of bound--state--QED which include the 
one--loop--vacuum polarization.

\section{The renormalized expression of S(VP)E}

As we have seen in the previous section the counter term vanishes 
and therefore it is only necessary to subtract the known Uehling contribution 
of the S(VP)E diagram. The renormalized expression of the energy correction of higher 
order is shown in Fig. \ref{fig13} and the corresponding energy shift reads:
\bea
\bigtriangleup E_n^{\rm S(VP)E} &=& e^4 \int\limits_{-\infty}^{\infty}
\frac{d E_1}{2\pi} \int\limits_{-\infty}^{\infty}\frac{d E_2}{2\pi} 
\int\frac{d^3 {\mbox{\boldmath $r$}_1}}{(2\pi)^3}\int\frac{d^3 {\mbox{\boldmath $r$}_2}}{(2\pi)^3}
\int\frac{d^3 {\mbox{\boldmath $r$}_3}}{(2\pi)^3}\int\frac{d^3 {\mbox{\boldmath $r$}_4}}{(2\pi)^3}\nonumber\\
\nonumber\\
&&\hspace{-2.0cm}\times\varphi_n^{\dagger} ({\mbox{\boldmath $r$}_1}) \alpha_{\mu} S_F (E_n-E_1,{\mbox{\boldmath $r$}_1},{\mbox{\boldmath $r$}_2})
\alpha_{\sigma} \varphi_n ({\mbox{\boldmath $r$}_2})
D^{\mu \nu} (E_1,{\mbox{\boldmath $r$}_1},{\mbox{\boldmath $r$}_3}) D^{\rho \sigma} (E_1,{\mbox{\boldmath $r$}_4},{\mbox{\boldmath $r$}_2})\nonumber\\
\nonumber\\
&&\hspace{-2.0cm}\times{\rm Tr} \left[ \alpha_{\nu} S_F (E_1+E_2,{\mbox{\boldmath $r$}_3},{\mbox{\boldmath $r$}_4})
\alpha_{\rho} S_F (E_2,{\mbox{\boldmath $r$}_4},{\mbox{\boldmath $r$}_3})\right]\;.
\label{eq_section1_65}
\eea
The mathematical expression for the subtracted Uehling--diagram can be obtained 
from Eq. (\ref{eq_section1_65}) by the replacement of the bound--electron propagator 
under the trace by the free--electron propagator.
\section{Summary}

In this article, using the Bogoljubov--Parasiuk--Hepp--Zimmermann 
(BPHZ)--renor\-malization method, we derived renormalized 
expressions for the energy shift of higher--order for the last two unknown 
second--order diagrams of bound state QED, the two--photon vacuum polarization 
VPVP and self energy vacuum polarization S(VP)E, respectively.
It was shown in some detail that counter terms of the outer renormalization 
for the VPVP diagram are cancelled against each other. This result simplifies 
signifiantly numerical evaluations of this diagram.\\ 
In a second part of this paper it has been proven for 
a wide class of diagrams for bound--state--QED which contains the 
one--loop vacuum polarization VP that the counter term of the
fourth--rank vacuum polarization tensor vanishes since  
the sum over the quantum number $\kappa$ has been performed 
until a finite $\kappa_{max}$ and as the last step in the spherical expansion 
of the electron propagator.  
It has been 
discussed in some detail that gauge invariance is not broken in such cases. 
The disappearance of the counter term simplify essentially the 
evaluation of such diagrams which contain the diagram VP. The 
investigations presented here are to be considered as generalization of corresponding 
investigations presented in \cite{section2_2} and \cite{section2_4}. 
We also mention that the numerical evaluation of VPVP and S(VP)E for 
hydrogenlike ions has to be consider as important step in the future   
to reduce the uncertainty in Lamb shift predictions and still is under 
consideration. 
Simultaneously, the results of this paper establish the use of the 
BPHZ--approach in bound state QED. 

\section*{Appendix}
 
The bound--electron propagator is given by  
\be
S_F (E,{\mbox{\boldmath $r$}_1},{\mbox{\boldmath $r$}_2}) = \sum \limits_{n,\kappa,\mu} 
\frac{\varphi_{n,\kappa,\mu} ({\mbox{\boldmath $r$}_1}) \overline{\varphi}_{n,\kappa,\mu} ({\mbox{\boldmath $r$}_2})}
{E - E_{n,\kappa} (1 - i \epsilon)}\;,
\label{section3_35}
\ee
where $\varphi$ are solutions of the Dirac equation with Coulomb potential:
\be
\left( -i \mbox{\boldmath$\alpha$}\nabla + \beta 
+ V ({\mbox{\boldmath $r$}}) \right) \varphi ({\mbox{\boldmath $r$}}) = E \varphi ({\mbox{\boldmath $r$}}) \;.
\label{section3_40}
\ee
The free--electron propagator of Eq. (\ref{eq_section1_15}) can be rewritten 
in a similar form as the bound--electron propagator: 
\bea
S_F^0 (E,{\mbox{\boldmath $r$}_1} - {\mbox{\boldmath $r$}_2}) &=& \int \limits_0^{\infty} d p 
\sum \limits_{\kappa,\mu} \sum \limits_{s=1}^2
\frac{\psi_{s,\kappa,\mu} ({\mbox{\boldmath $r$}_1}) \overline{\psi}_{s,\kappa,\mu} ({\mbox{\boldmath $r$}_2})}
{E - E_{s,p} (1 - i \epsilon)}\;,
\label{eq_section3_42}
\eea
where the index $s$ signifies the positive and negative energy states of a free 
electron
\bea
E_{s,p} = \pm \sqrt{m^2+\mbox{\boldmath $p$}^2}\;,\quad (s=1,2)\;,
\label{eq_section3_43}
\eea
and $\Psi$ are the solutions of the free Dirac equation:
\be
\left( -i \mbox{\boldmath$\alpha$}\nabla + \beta \right) 
\Psi ({\mbox{\boldmath $r$}}) = E\,\Psi ({\mbox{\boldmath $r$}}) \; .
\label{eq_section3_44}
\ee
The free--electron propagator in spherical expansion can be written as 
\cite{lit_anhang_5}:
\be
\begin{array}[c]{l}
\displaystyle
S_F^0 (E, {\mbox{\boldmath $r$}_1}-{\mbox{\boldmath $r$}_2}) = \sum \limits_{\kappa}
\left ( 
\begin{array}[c]{l}
G^{11}_{0,\kappa} (E,r_1,r_2) \; \pi^{11}_{\kappa} \quad \quad 
G^{12}_{0,\kappa} (E,r_1,r_2) \; \pi^{12}_{\kappa} \\
\\
\displaystyle
G^{21}_{0,\kappa} (E,r_1,r_2) \; \pi^{21}_{\kappa} \quad \quad 
G^{22}_{0,\kappa} (E,r_1,r_2) \; \pi^{22}_{\kappa}
\end{array}
\right )\;,
\end{array}
\label{eq_free_40}
\ee
with the radial components (for $r_1>r_2$)
\bea
G^{11}_{0,\kappa} (E, r_1,r_2) =
- (iE+1) \sqrt{E^2+1}\; j_{|\kappa+1/2|-1/2} (ir_1\sqrt{E^2+1})\; 
h_{|\kappa+1/2|-1/2}^{(1)} (ir_2\sqrt{E^2+1}) \;,
\nonumber\\
\nonumber\\
G^{12}_{0,\kappa} (E, r_1,r_2) = -i (E^2+1) \;{\rm sign} (\kappa) \;
j_{|\kappa+1/2|-1/2} (ir_1 \sqrt{E^2+1}) \;
h_{|\kappa-1/2|-1/2}^{(1)} (ir_2 \sqrt{E^2+1}) \;,
\nonumber\\
\nonumber\\
G^{21}_{0,\kappa} (E, r_1,r_2) = -i (E^2+1)\; {\rm sign} (\kappa) \;
j_{|\kappa-1/2|-1/2} (ir_1\sqrt{E^2}+1) \;
h_{|\kappa+1/2|-1/2}^{(1)} (ir_2\sqrt{E^2+1}) \;,
\nonumber\\
\nonumber\\
G^{22}_{0,\kappa} (E, r_1,r_2) = 
- (iE-1) \sqrt{E^2+1}\; j_{|\kappa-1/2|-1/2} (ir_1\sqrt{E^2+1}) \;
h_{|\kappa-1/2|-1/2}^{(1)} (ir_2\sqrt{E^2+1}) \;.
\nonumber
\eea
\vspace{-1.0cm}
\bea
\label{eq_free_45}
\eea
For the case $r_2>r_1$ one simply 
can employ the symmetry relations:
\bea
G^{11}_{0,\kappa} (E, r_1,r_2) &=& G^{11}_{0,\kappa} (E, r_2,r_1) \;, \nonumber\\
G^{12}_{0,\kappa} (E, r_1,r_2) &=& G^{21}_{0,\kappa} (E, r_2,r_1) \;, \nonumber\\
G^{21}_{0,\kappa} (E, r_1,r_2) &=& G^{12}_{0,\kappa} (E, r_2,r_1) \;, \nonumber\\
G^{22}_{0,\kappa} (E, r_1,r_2) &=& G^{22}_{0,\kappa} (E, r_2,r_1) \;.
\label{eq_free_46}
\eea
In Eq. (\ref{eq_free_40}) $\pi_{\kappa}^{ij}$ denote the spin--angualar functions 
which are given by:
\bea
&&{\pi}^{11}_{\kappa} = \sum \limits_{\mu}  {\chi}_{\kappa}^{\mu} (\theta_1,\varphi_1)\;
{\chi}^{\mu\;\dagger}_{\kappa} (\theta_2,\varphi_2) \;; \quad\;\;\;
{\pi}^{12}_{\kappa} = \sum \limits_{\mu}  {\chi}_{\kappa}^{\mu} (\theta_1,\varphi_1) \;
{\chi}^{\mu\;\dagger}_{-\kappa} (\theta_2,\varphi_2)\;; \nonumber\\
\nonumber\\
&& {\pi}^{21}_{\kappa} = \sum \limits_{\mu}  {\chi}_{-\kappa}^{\mu} (\theta_1,\varphi_1)\;
 {\chi}^{\mu\;\dagger}_{\kappa} (\theta_2,\varphi_2) \;; \quad
 {\pi}^{22}_{\kappa} = \sum \limits_{\mu}  {\chi}_{-\kappa}^{\mu} (\theta_1,\varphi_1)\;
 {\chi}^{\mu\;\dagger}_{-\kappa} (\theta_2,\varphi_2) \;.
\label{eq_2_1_140}
\eea
and $\chi_{\kappa}^{\mu}$ is defined as:
\be
\begin{array}[t]{l}
\displaystyle
\chi_{\kappa}^{\mu} (\theta, \phi) = \left ( \begin{array}[c]{l} 
\displaystyle
- \frac{\kappa}{|\kappa|} \sqrt {\frac{\kappa + \frac{1}{2} - \mu}{2 \kappa +1}}
\;Y_{|\kappa + \frac{1}{2}| - \frac{1}{2}, \mu - \frac{1}{2}} 
(\theta, \phi) \\
\\
\displaystyle
\sqrt {\frac{\kappa + \frac{1}{2} + \mu}{2 \kappa +1}}\;
Y_{|\kappa + \frac{1}{2}| - \frac{1}{2}, \mu + \frac{1}{2}}
(\theta, \phi)
\end{array}
\right)
\end{array}\;,
\label{eq_2_1_45}
\ee
where $Y_{|\kappa + \frac{1}{2}| - \frac{1}{2}, \mu \pm \frac{1}{2}} (\theta, \phi)$ 
are the spherical harmonics.
The photon propagator reads in Feynman gauge:
\be
D_{\mu \nu} (t_1-t_2,{\mbox{\boldmath $r$}_1}-{\mbox{\boldmath $r$}_2}) = - g_{\mu \nu} \int 
\limits_{-\infty}^{\infty} \frac{d E}{2 \pi} \int
\frac{d^3 \mbox{\boldmath $k$}}{(2 \pi)^3} \frac{e^{- i [E (t_1-t_2) - \mbox{\scriptsize\boldmath $k$} ({\mbox{\scriptsize\boldmath $r$}_1}-{\mbox{\scriptsize\boldmath $r$}_2})]}}
{E^2 - \mbox{\boldmath $k$}^2 + i \epsilon}\;.
\label{eq_2_1_50}
\ee

\section*{Acknowledgements}
Financial support from BMBF, DAAD, DFG and GSI is greatefully acknowledged.
The authors would like to thank Prof. L. N. Labzowsky, Dr. I. A. 
Goidenko, Dr. A. V. Nefiodov, Prof. V. M. Shabaev and Dr. V. A. Yerokhin for 
valuable discussions.

\newpage
\begin{figure}[!h]
\begin{center}
\includegraphics[scale=0.4]{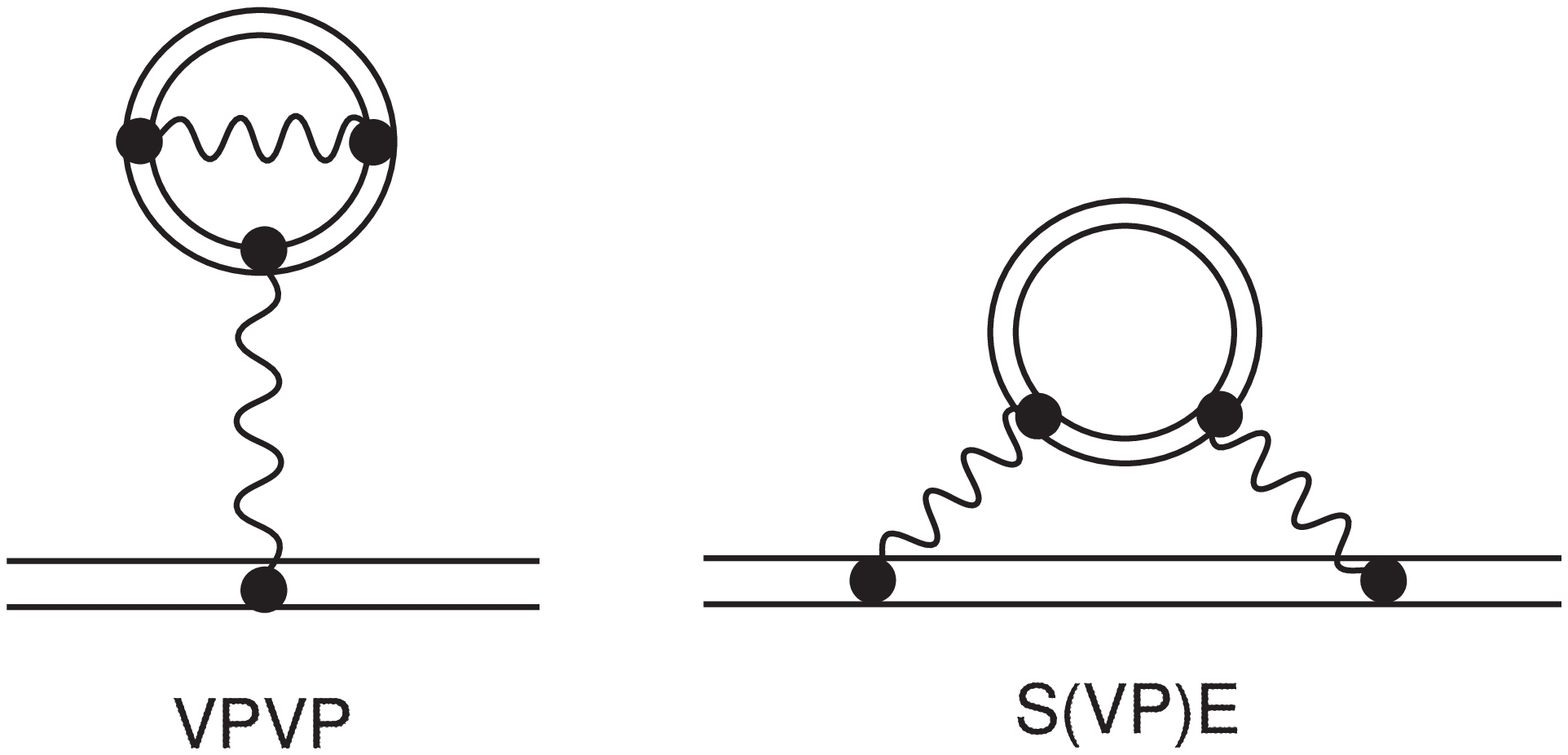}
\caption{The two--photon vacuum polarization (VPVP) and the effective 
self energy S(VP)E.}
\label{fig0}
\end{center}
\end{figure}

\begin{figure}[!h]
\begin{center}
\includegraphics[scale=1.0]{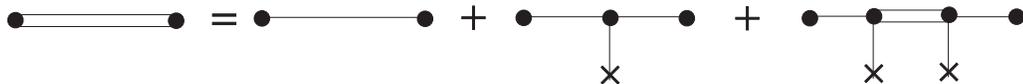}
\caption{Potential expansion of the
bound--electron propagator.}
\label{fig2}
\end{center}
\end{figure}

\begin{figure}[!h]
\begin{center}
\includegraphics[scale=0.4]{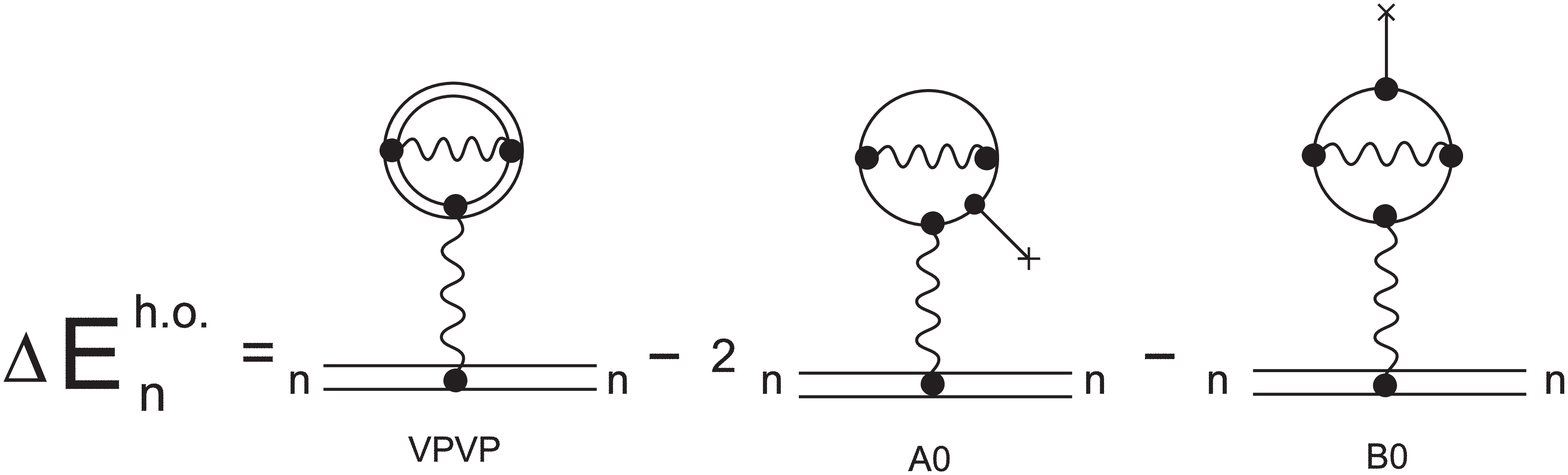}
\caption{Unrenormalized energy shift due to higher--order contribution in 
$(Z\alpha)$ of the VPVP diagram.}
\label{fig1}
\end{center}
\end{figure}

\begin{figure}[!h]
\begin{center}
\includegraphics[scale=0.3]{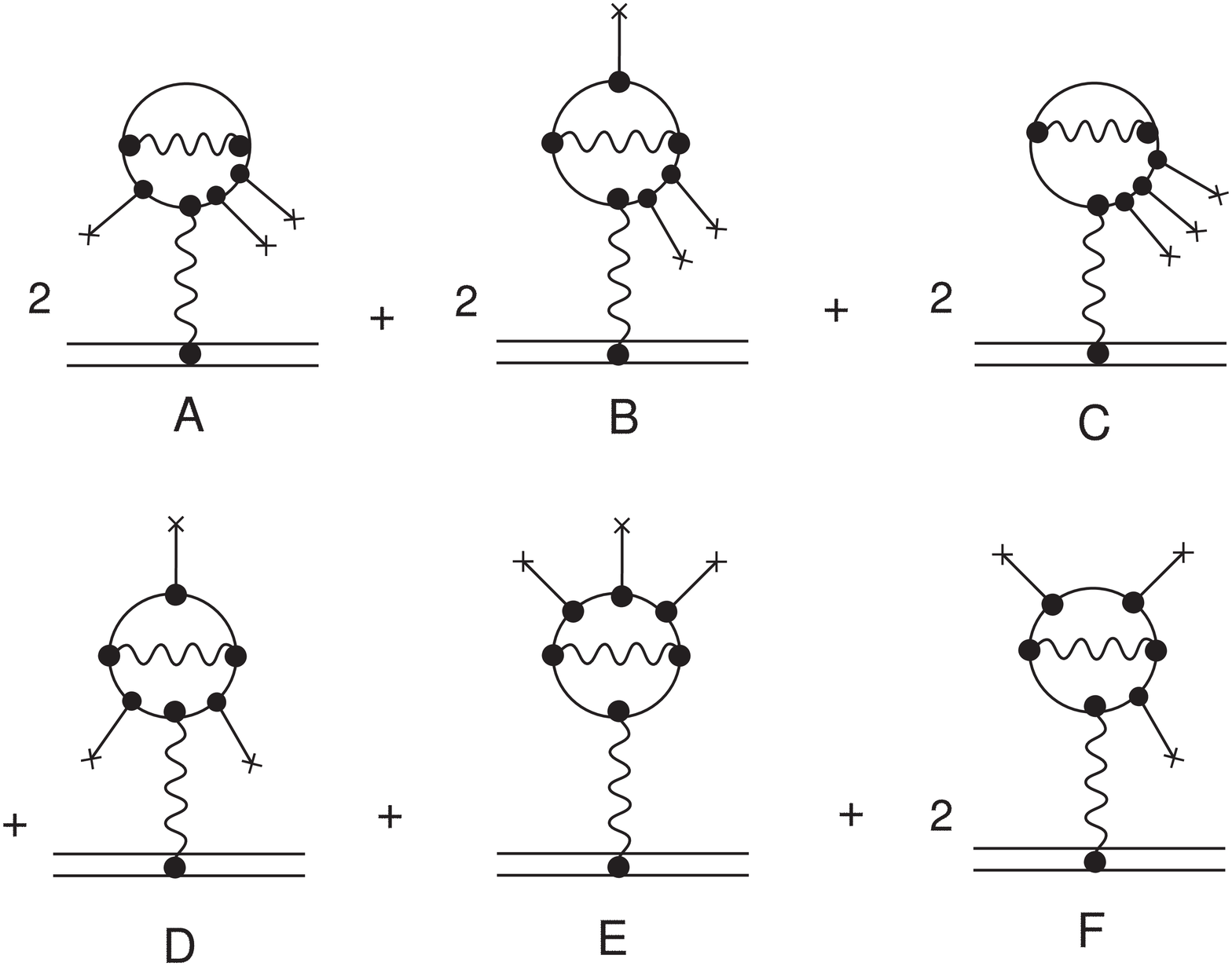}
\caption{Lowest--order terms of the potential expansion contributing
to the energy shift of Fig. \ref{fig1} with $\omega=0$.}
\label{fig3}
\end{center}
\end{figure}

\begin{figure}[!h]
\begin{center}
\includegraphics[scale=0.5]{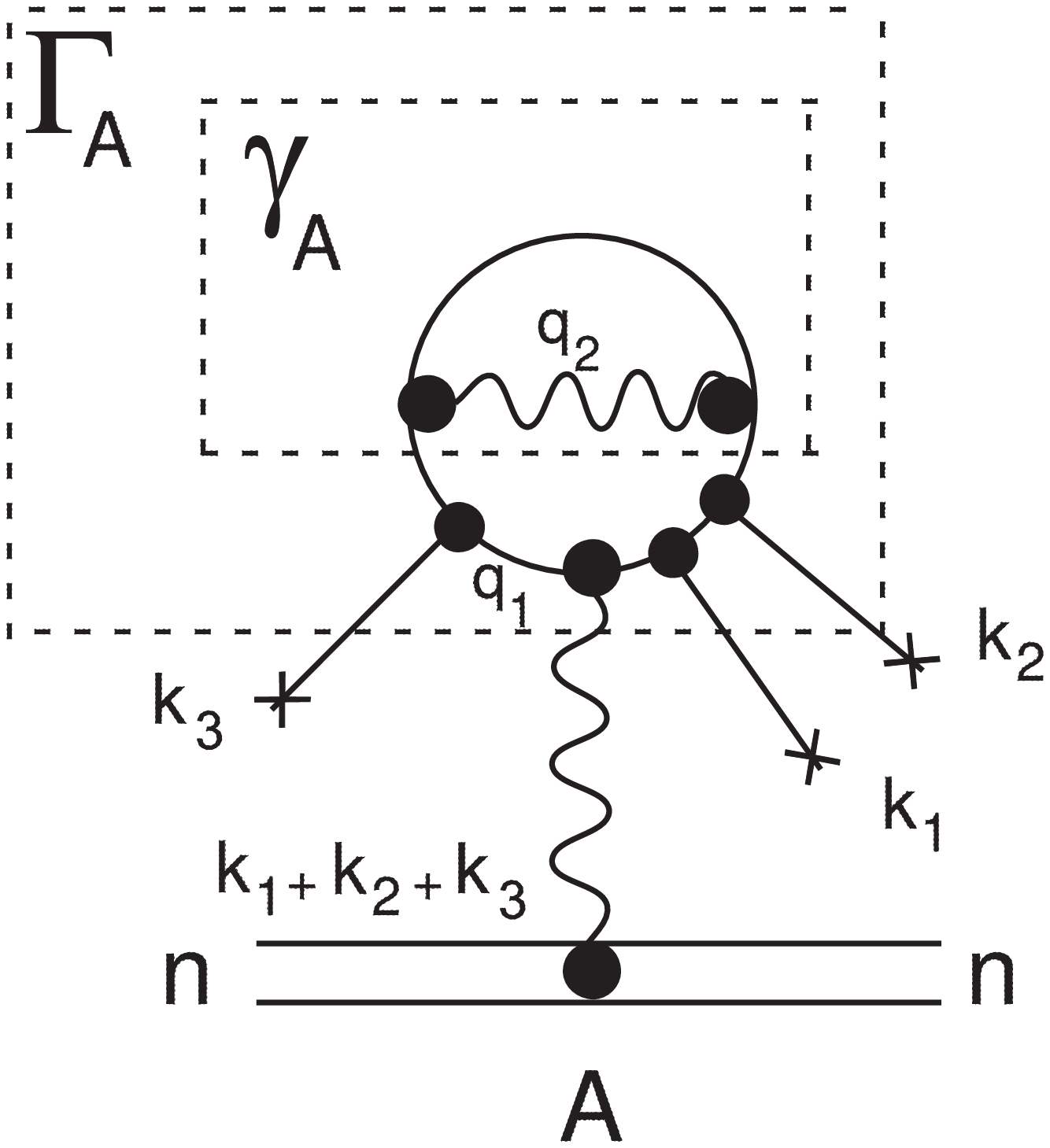}
\caption{Diagram $\Gamma_A$ with $\omega(\Gamma_A)=0$. $\Gamma_A$ 
contains a divergent subdiagram $\gamma_A$ (= one--loop self energy). 
n denotes a bound electron state.}
\label{fig4}
\end{center}
\end{figure}

\newpage

\begin{figure}[!h]
\begin{center}
\includegraphics[scale=0.3]{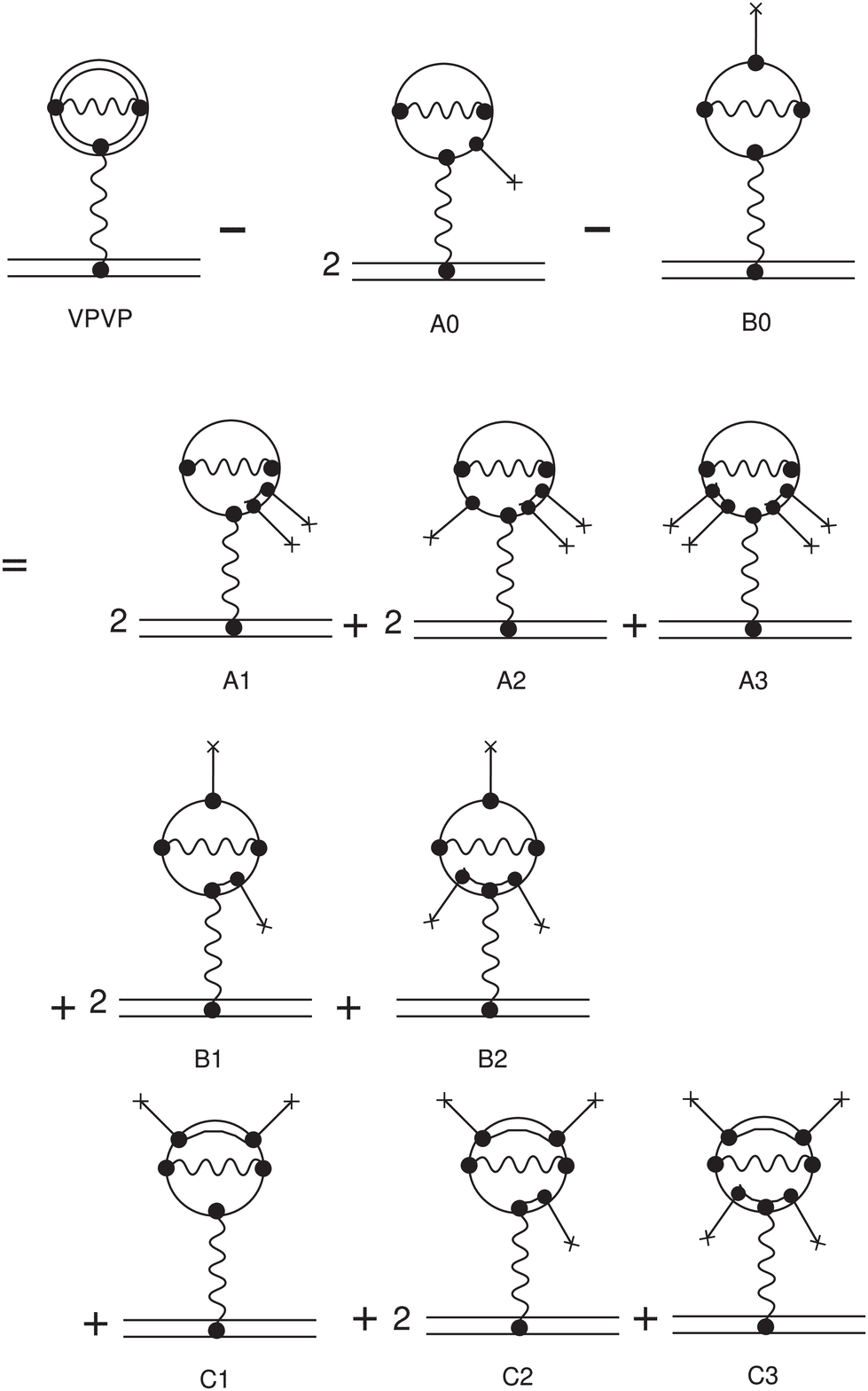}
\caption{The potential expansion of the unrenormalized expression 
Fig. \ref{fig1}.}
\label{fig5}
\end{center}
\end{figure}

\newpage

\begin{figure}[!h]
\begin{center}
\includegraphics[scale=0.25]{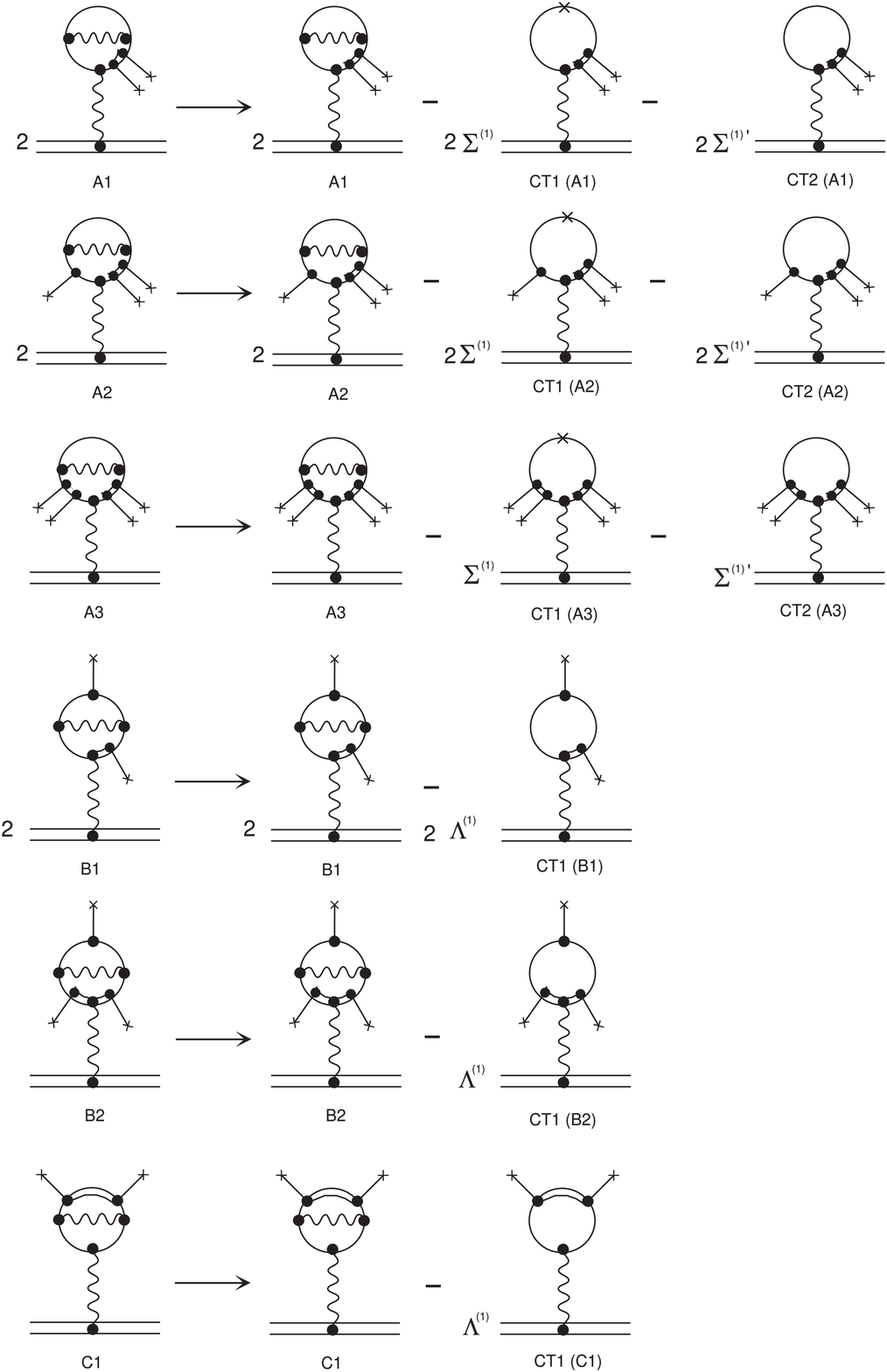}
\caption{The inner renormalization according to Eq. (\ref{section2_60})}
\label{fig6}
\end{center}
\end{figure}

\newpage

\begin{figure}[!h]
\begin{center}
\includegraphics[scale=0.35]{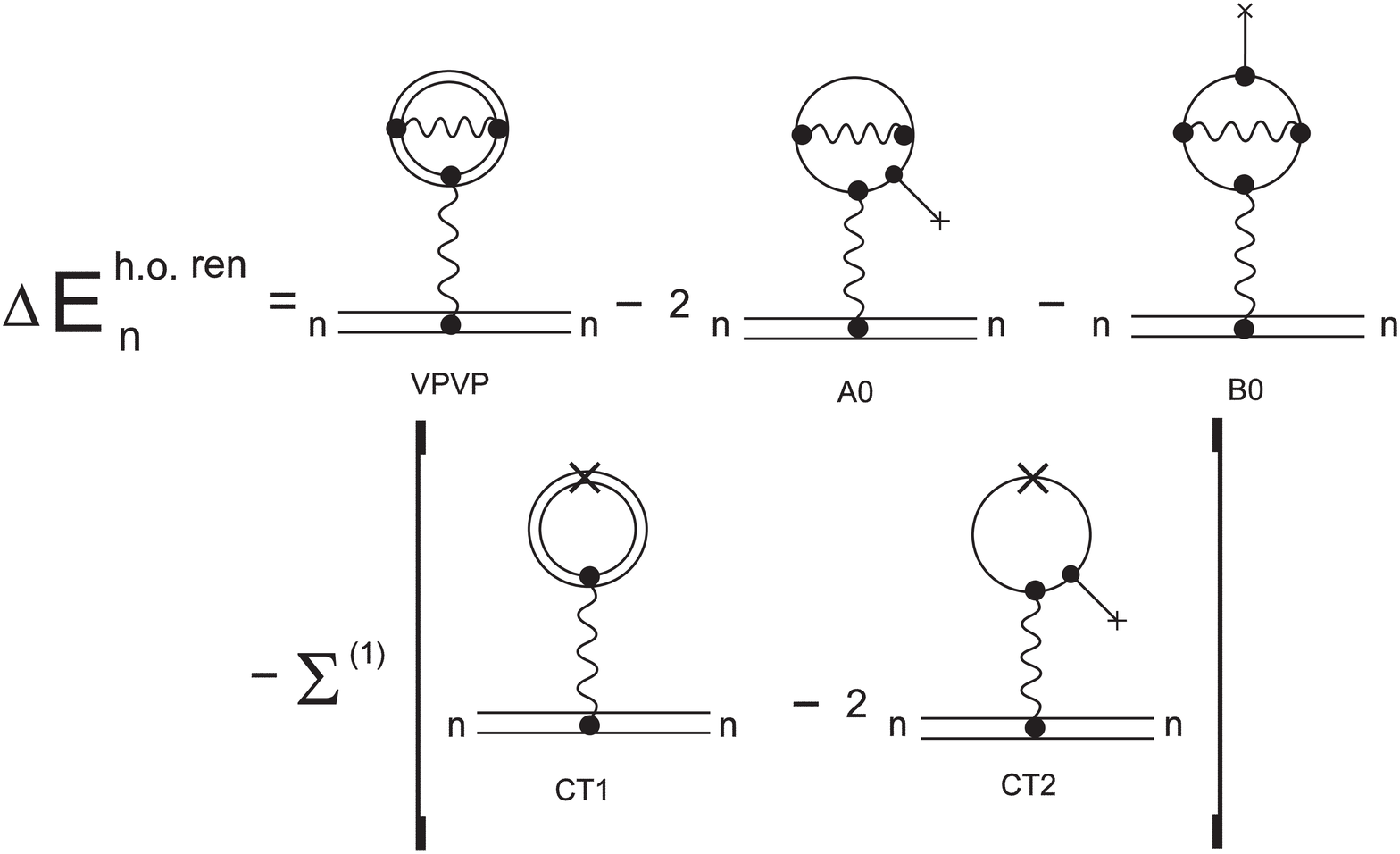}
\caption{This is the final expression for the renormalization of the VPVP 
diagram if one subtracts the K\"all\'en-Sabry--terms.}
\label{fig7}
\end{center}
\end{figure}

\newpage

\begin{figure}[!h]
\begin{center}
\includegraphics[scale=0.4]{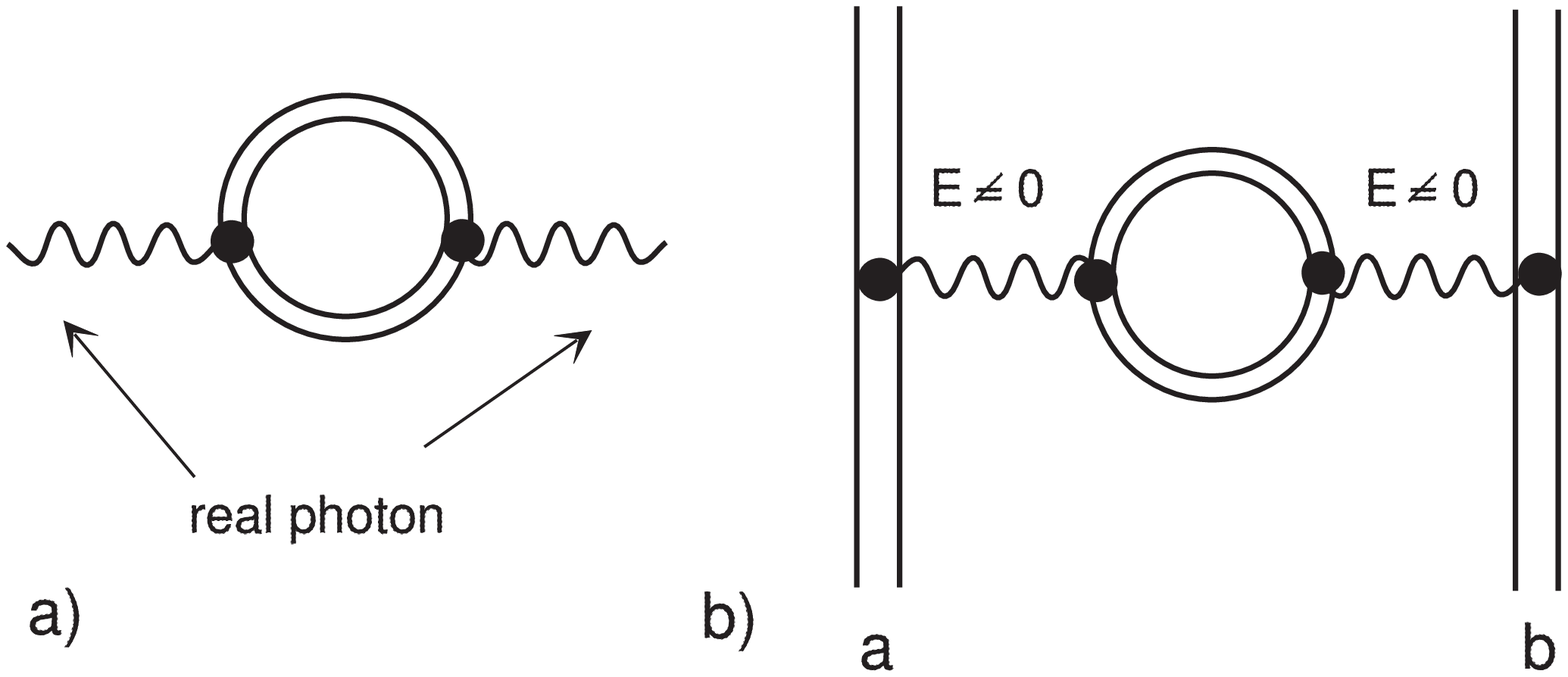}
\caption{The diagrams of Delbr\"uck scattering a) and energy splitting b).}
\label{fig8}
\end{center}
\end{figure}

\begin{figure}[!h]
\begin{center}
\includegraphics[scale=0.45]{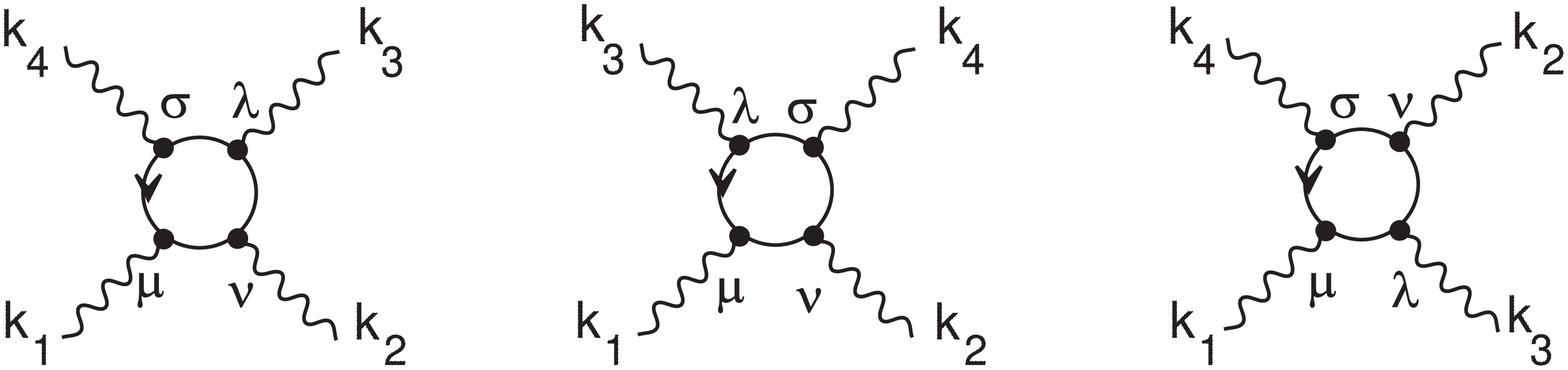}
\caption{The diagrams of light--light scattering.}
\label{fig9}
\end{center}
\end{figure}

\begin{figure}[!h]
\begin{center}
\includegraphics[scale=0.2]{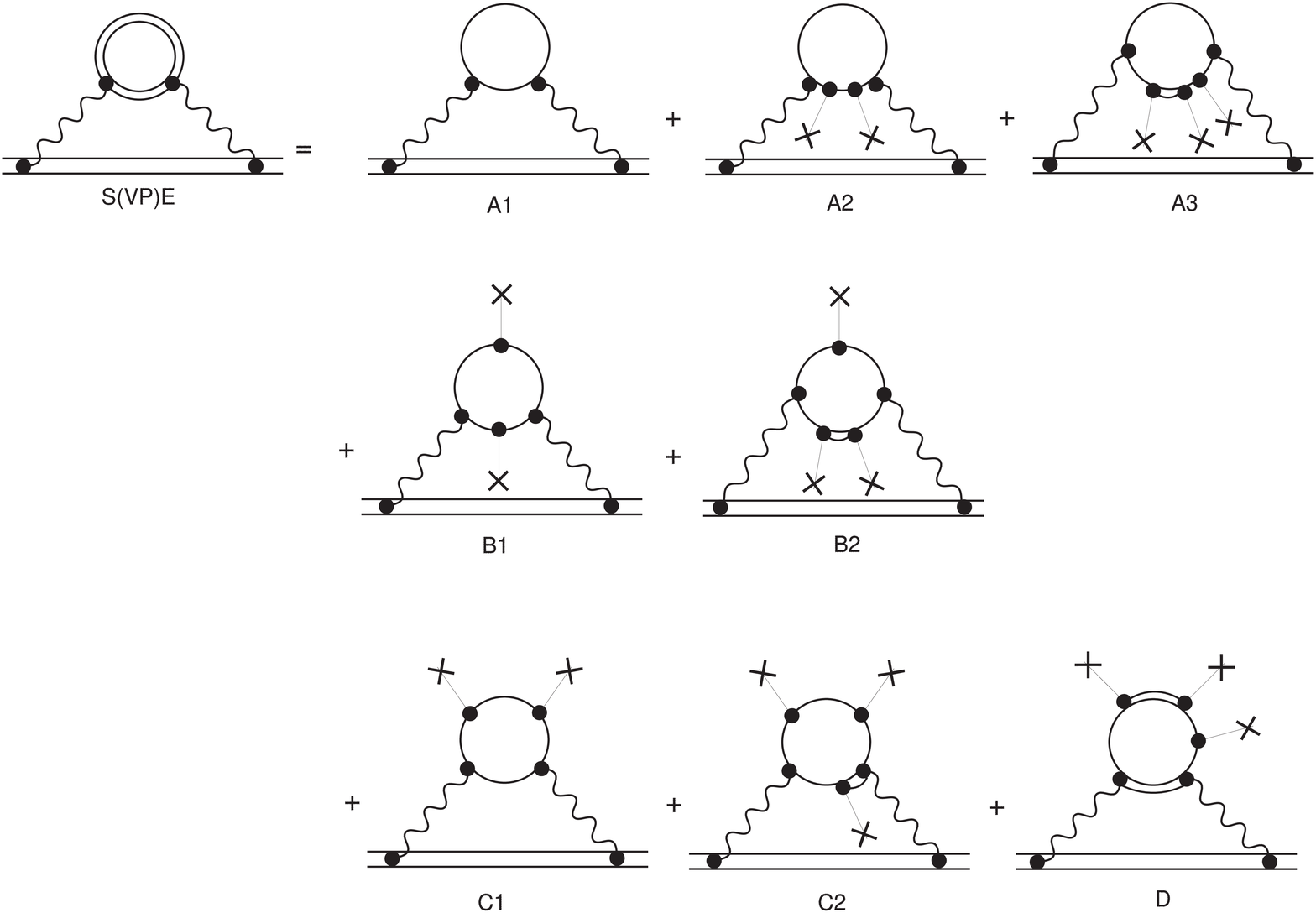}
\caption{The potential expansion of the S(VP)E diagram. The diagrams 
A3, B2, C2 and D are convergent. 
The diagrams A2, B1 and C1 are separately divergent but their sum is 
convergent but not gauge invariant. Diagram A1 is the so--called 
Uehling contribution of the S(VP)E.}
\label{fig11}
\end{center}
\end{figure}

\begin{figure}[!h]
\begin{center}
\includegraphics[scale=0.35]{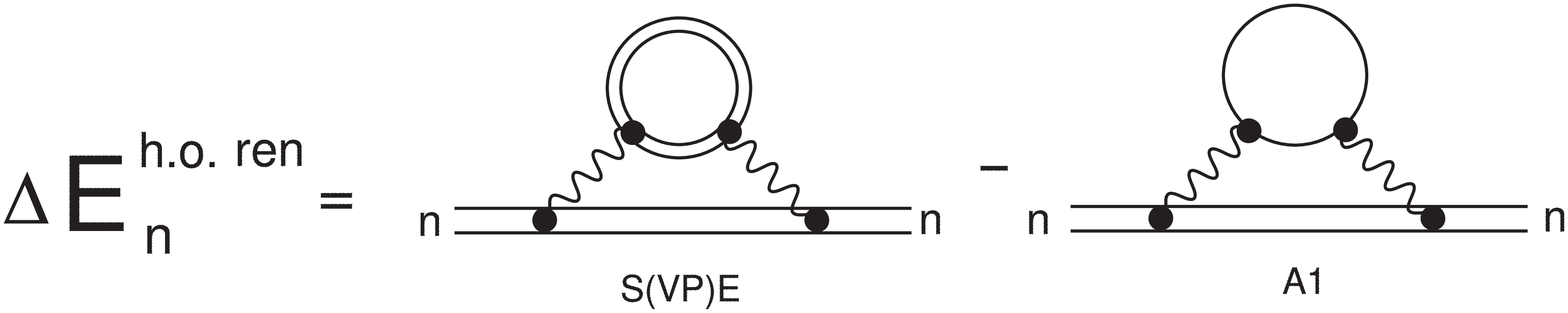}
\caption{The renormalized expression of energy shift of higher--order of 
the self energy vacuum polarization.}
\label{fig13}
\end{center}
\end{figure}

\end{document}